\documentclass[useAMS,usenatbib,referee]{mn2e}
\usepackage{multirow}
\usepackage{graphicx,color}
\newcommand{\etal }{{et al.} }
\newcommand{\msun}{\thinspace M_\odot}

\newcommand{\vect}[1]{\mbox{\boldmath$#1$}}
\def\lesssim{\mathrel{\hbox{\rlap{\hbox{\lower4pt\hbox{$\sim$}}}\hbox{$<$}}}}
\def\gtrsim{\mathrel{\hbox{\rlap{\hbox{\lower4pt\hbox{$\sim$}}}\hbox{$>$}}}}
\newcommand{\cm}{\,{\rm cm}^{-3} }

\newcommand{\mdot}{M_\odot\,{\rm yr}^{-1} }

\newcommand{\dfrac}[2]{{\displaystyle \frac{#1}{#2}} }

\title[Massive Outflows]{Massive Outflows Driven by Magnetic Effects in Star Forming Clouds with High Mass Accretion Rates}
\author[Y. ~Matsushita,  \etal]
  { Yuko Matsushita$^{1}$, Masahiro N. Machida$^{1}$, Yuya Sakurai$^{2}$ and Takashi Hosokawa$^{2,3}$\thanks{E-mail:yuko.matsushita.272@s.kyushu-u.ac.jp (YM),  machida.masahiro.018@m.kyushu-u.ac.jp (MNM)}  \\
$^{1}$ Department of Earth and Planetary Sciences, Faculty of Sciences, Kyushu University, Fukuoka 819-0395, Japan\\
$^{2}$ Department of Physics, The University of Tokyo, Tokyo 113-0033, Japan\\
$^{3}$Department of Physics, Graduate School of Science, Kyoto University, Sakyo-ku, Kyoto 606-8502, Japan
}

\begin{document}
\maketitle
\begin{abstract}
The relation between the mass accretion rate onto the circumstellar disc and the rate of mass ejection by magnetically driven winds is investigated using three-dimensional magnetohydrodynamics simulations. 
Using a spherical cloud core with a varying ratio of thermal to gravitational energy, which determines the mass accretion rate onto the disc, to define the initial conditions, the outflow propagation for approximately $10^4$\,yr after protostar formation is then calculated for several cloud cores.
The mass ejection rate and accretion rate are comparable only when the magnetic energy of the initial cloud core is comparable to the gravitational energy. 
Consequently, in strongly magnetised clouds a higher mass accretion rate naturally produces both massive protostars and massive outflows. 
The simulated outflow mass, momentum, kinetic energy and momentum flux agree well with observations, indicating  that massive stars form through the same mechanism as low-mass stars but require a significantly strong magnetic field to launch massive outflows. 
\end{abstract}

\clearpage
\section{Introduction}
\label{sec:intro}
Stellar masses are distributed over four orders of magnitude from $\sim0.01 \msun$ to $\gtrsim 100\msun$.
Observations of the star forming regions in our galaxy reveal that the stellar initial mass function (IMF) peaks at  $\sim0.5\msun$, indicating that the galaxy is mostly populated by low-mass stars \citep{kroupa02,chabrier03}. 
The low-mass star formation process has been thoroughly investigated and supported by numerous theoretical works \citep[e.g.][]{larson69,masunaga00}.
Conversely, because massive stars (with masses $\gtrsim 8\msun$) are much rarer than low-mass stars \citep{kroupa02} and form in regions far from the Sun, their formation process is less clearly understood. 
However, massive stars can greatly disturb the star forming environment through stellar radiative feedback, injecting kinetic and thermal energy as well as polluted gas into interstellar space through powerful stellar winds and supernova explosions \citep{zinnecker07,beuther07,tan14}. 
Thus, despite being relatively low in number, massive stars significantly affect the evolution of the galaxy, and clarifying their formation process is an important aspect to astrophysical study.

Protostellar outflow offers useful clues for elucidating the star formation process. 
Such outflows are ubiquitously observed around young low-mass protostars \citep{bontemps96,cabrit92,wu04,hatchell07,arce10}, and their driving mechanism, which has been investigated in many theoretical studies, can be briefly summarized as follows.
Once a prestellar cloud core begins to collapse, gravitational energy is converted into magnetic and kinetic energy that  contribute to the driving outflow either through magnetic pressure or magneto-centrifugal forces around the protostar or circumstellar disc \citep{blandford82,uchida85,pudritz86,tomisaka02,lynden-Bell03,machida08}. 
Protostellar outflow significantly influences the low-mass star formation process through the transport of excess angular momentum from the protostar forming region or circumstellar disc into interstellar space, enabling the infalling gas to reach the protostar. 
As such, it promotes the mass accretion and protostellar growth during the early star formation phase \citep{pudritz83,konigl89, wardle93, tomisaka02}. 
Protostellar outflow also plays an important role in the low-mass star formation process by sweeping up much of the infalling gas in the collapsing cloud and expelling it into interstellar space. 
Thus, protostellar outflow driven by a low-mass protostar is thought to control star formation efficiency and determine the resultant stellar mass in the cloud \citep{nakano95,matzner00,machida13}.

Recent developments in observational techniques and technologies have enabled the observation of massive star forming regions at high spatial resolutions and have led to the confirmation of many massive outflows believed to be driven by massive protostars \citep{beuther02, wu04, wu05, zhang05, lopes09,villiers14,villiers15,maud15}.
Early observations hinted that outflows driven by massive protostars have wider opening angles than those driven by low-mass protostars; consequently, massive and low-mass outflows were originally regarded as different phenomena \citep{richer00,ridge01}. 
However, as the observations accumulated, it was confirmed that the structure and shape of the outflows driven by massive and low-mass protostars are very similar \citep{qiu09, beltran11, beuther11, palau13, villiers14, caarrasco15}. 
Observations have also shown that the physical quantities of outflows driven by massive protostars are strongly correlated with protostellar luminosity or mass. 
In particular, the mass, momentum, momentum flux and kinematic energy of outflows increase with increasing protostellar mass. 
Consequently, a massive protostar has a massive protostellar outflow. 
The correlation between the physical quantities of the outflow and the bolometric luminosity or protostellar mass is seen over a very wide mass range from low- to very high-mass protostars \citep{beuther02, wu05, zhang05, villiers14,maud15}. 
This implies that massive outflows are scaled-up versions of  low-mass outflows. 
Although massive star forming regions are distant and it is not possible to resolve massive protostars themselves, the physical quantities of their massive outflows can be observationally determined because the outflows ($\gtrsim 0.1-1$\,pc) have much larger dimensions than the protostars ($\lesssim 0.1$\,AU). 
Thus, these massive outflows can provide important clues for understanding the formation processes of massive stars.

The relation between the core mass function (CMF) and IMF may provide further insights into star formation processes. 
The similarity between CMF and IMF holds over a wide range of stellar masses \citep{motte98,enoch06,nutter07,alves07,konyves10,andre10} and indicates that only $\sim30-50\%$ of the prestellar cloud mass can be converted into stellar mass (i.e. the star formation efficiency is $\varepsilon\sim30-50\%$). 
In low-mass star formation, the remaining mass is considered to be ejected from the parent cloud into interstellar space by protostellar outflow. 
The similarity between CMF and IMF also holds in massive stars and clouds, indicating that massive stars also form from a fraction of the parent cloud mass \citep{nakano95,matzner00}. 
As low-mass and massive cloud cores have almost the same star formation efficiency, they likely form by the same fundamental mechanism, which can be surmise to function as follows: a massive cloud core collapses into a protostar, which then grows by mass accretion from the infalling envelope (the so-called core accretion scenario, e.g. \citealt{tan14}). 
In this case, massive star formation can be most simply characterised by two parameters: the initial cloud mass and the mass accretion rate. 
A massive cloud core provides the necessary mass reservoir for massive star formation and a large accretion rate enables the rapid formation of a massive star. The latter conditions are naturally satisfied in gravitationally highly unstable cloud cores \citep{bergin07}.

In low-mass star formation, the protostellar outflow emerges immediately after (or before) protostar formation and sustains itself during the mass accretion stage from the gravitational energy released by the infalling gas \citep{machida12}.
A similar evolutionary trend is expected for the massive star formation process, with  the outflow from massive protostars also determining (at least partially) the resultant stellar mass. 
Note that radiative effects are also expected to significantly contribute to the determination of the final stellar mass (for details, see \S\ref{sec:feedback}).

In the low-mass star formation process, outflow is driven by magnetic effects, which also collimate the outflow via a strong toroidal field and hoop stress \citep{pudritz07}. 
Prestellar clouds in low-mass star forming regions have strong magnetic fields, with magnetic energies comparable to their gravitational energies \citep{crutcher99,troland08,Crutcher10}. 
Such fields are necessary for driving protostellar outflow \citep{hennebelle11,seifried11,machida13}.

Recently, strong magnetic fields have also been observed in massive star forming regions \citep{falgarone08}. 
Such magnetic fields significantly affect the dynamical evolution of massive clumps and massive star formation \citep{girart09,tan14,li15,tan16}. 
\citet{zhang14} showed that, at cloud core scales of $\sim0.01-0.1$\,pc, the magnetic field (rather than turbulence) primarily controls  massive star formation. 
This result is not surprising because the cloud collapses follows the magnetic field lines after turbulence dissipation, even in initially highly turbulent clouds.
When the parent cloud of the massive star has a strong field, star formation can be controlled by the protostellar outflow, as is true in low-mass star formation (see \S\ref{sec:feedback} for radiation feedback effect).
Note, however, that when the turbulence is super-Alfv\'enic, it may sustain for a long time \citep[e.g.][]{burkhart15}, in which case it may greatly affect the massive star formation process. 
We need further observations of massive star forming regions to determine which effect primarily controls massive star formation.

Based on the core accretion scenario, many theoretical studies have investigated the outflow driven by magnetic effects in the low-mass star formation process \citep[e.g.][]{tomisaka02,machida08,hennebelle08,price12,tomida13,machida13,machida14,bate14,tsukamoto15,tomida15}. 
By contrast, magnetically driven outflow in massive star formation has been investigated in only a few studies \citep{peters11,commercon11, hennebelle11, seifried12,kuiper15}. 
Moreover, these studies have been limited to cloud cores with a specific cloud size, mass and accretion rate. Thus, the authors of these studies made no direct connections between low-mass (or intermediate-mass) and massive star formation processes.

In this study, we aim to derive the common mechanism of star formation processes using magnetohydrodynamics (MHD) simulations. 
We investigate the high-mass (or massive) star formation process and clarify the fundamental star formation process applicable to low/intermediate-mass and star formation. 
Under simple settings, we calculate the evolution of collapsing clouds at varying masses and stabilities (ratio of thermal to gravitational energy of the initial cloud), which control the mass accretion rate and characterise the resultant stellar mass (\S\ref{sec:accretionrate}). 
As the outflow is closely related to the gravitational collapse and accretion process (the primary mechanisms in the core accretion scenario), we closely investigate the outflow properties and compare them with observations to validate the core accretion scenario.
Note that, as described in \S2, we set the prestellar cloud mass to $\ge 32\msun$ and calculate the cloud evolution until the protostellar mass reaches $\gtrsim 2\msun$. 
Thus, this study focuses on intermediate or high-mass star formation. 
When discussing the star formation process from low- to high-mass stars, we refer to our previous study \citep{machida13} in addition to the results of this study. 

The remainder of this paper is structured as follows. \S2 describes the basic equations and numerical settings of out model and \S3 presents the simulation results. 
In \S4, we compare our results with observations and previous works. 
In \S5, we summarise the relation between magnetically driven outflow and the star formation process.

\section{Numerical Settings}
\label{sec:model}
We evolve gravitationally unstable cloud cores and outflow propagation using resistive MHD simulations that impose  a sink around the centre of the computational domain following protostar formation (\S\ref{sec:method}). 
In this way, it is possible to uniquely determine the protostellar mass and the mass accretion rate onto the protostar. 
However, because we ignore protostellar radiative feedback, which should affect the circumstellar environment, we evolve the system only up to a protostellar mass of $\sim30\msun$. 
During the corresponding period, the protostellar feedback is less significant than that at later stages \citep[e.g.][]{krumholz07,krumholz09}. 
The effect of radiative feedback is discussed in \S\ref{sec:feedback}.

\subsection{Basic Equations}
\label{sec:basic}
To evolve the star forming cloud, we use the resistive MHD nested grid method.
The numerical methods and initial settings are nearly identical to those described in our previous studies \citep{machida11c,machida12,machida13,machida14a,machida14}. 
The resistive MHD equations are given by Eqs. (1)--(4) in \cite{machida12}.
Instead of solving the energy equation or radiative transfer, we use a simple
barotropic equation of state:
\begin{equation} 
P =  c_{s,0}^2\, \rho \left[ 1+ \left(\dfrac{\rho}{\rho_{\rm cri}}\right)^{2/5} \right],
\label{eq:eos}
\end{equation}
where $c_{s,0}$ is the speed of sound in the initial cloud and $ \rho_{\rm cri}$ is the critical density.
We set $\rho_{\rm cri}=10^4\, \rho_{\rm c,0}$, where $\rho_{c,0}$ is the central density of the initial cloud.
The gas behaves isothermally when $\rho<\rho_{\rm cri}$ and contracts adiabatically when $\rho>\rho_{\rm cri}$.
The resistivity $\eta$ is that used in our previous works \citep{machida07,machida11b,machida13,machida14a}:
\begin{eqnarray}
\eta & = & C_{\eta} \, \eta_{\rm fid}, \nonumber \\
     & = & C_{\eta} \, 1.3\times 10^6 \left(\dfrac{T}{10 {\rm K}} \right)^{1/2} \left( \dfrac{n}{{\rm cm}^{-3}} \right),\label{eq:etadef}
\end{eqnarray}
where $\eta_{\rm fid}$ is determined by solving the chemical reactions of various neutral and charged particles \citep{nakano80,nakano86,nakano02}.
To investigate the effect of the dependence of magnetic diffusivity on the calculation results, we change the diffusivity through the parameter $C_\eta$, where $C_\eta=1$ for fiducial models and $C_{\eta}=0.01-100$ in various other models (for details, see \S\ref{sec:eta}).

\subsection{Initial Condition}
\label{sec:initial}
A spherically symmetric cloud with a Bonnor--Ebert density profile is adopted as the initial state. 
We set the initial cloud radius as twice the critical Bonner--Ebert radius  \citep{ebert55, bonnor56}. 
The Bonner-Ebert density profile is uniquely determined by the central density and isothermal temperature. 
To convert the non-dimensional parameters of the initial cloud into dimensional values, we assume a central density of $\rho_{\rm c,0}=3.8\times 10^{-19}$\,g\,cm$^{-3}$ (or number density $n_{c,0}=10^5\cm$) and an isothermal temperature of $T_{\rm iso}=40$\,K. 
With these parameters, the initial cloud has a dimensional radius and mass of $R_{\rm cl}=0.28$\,pc and $M_{\rm cl}=23\msun$, respectively. 
After constructing the dimensional cloud with the Bonnor--Ebert density profile, we increase the cloud density by a factor of $f$. 
The parameter $f$ controls the cloud mass and initial cloud stability $\alpha_0$, where $\alpha_0$ is the ratio of thermal to gravitational energy. 
When the cloud density is multiplied by $f$, the mass changes as $M_{\rm cl}=23\msun f$, and $\alpha_0$ also changes. 
Because the mass accretion rate is considered to be proportional to $\propto \alpha_0^{-3/2}$, the mass accretion rate is large in an initially thermally unstable cloud with larger $f$ or smaller $\alpha_0$.
\footnote{
The mass accretion rate can be described as $\dot{M} \propto M t_{\rm ff}^{-1}$, where $M$ is the prestellar cloud mass and $t_{\rm ff}$ is the freefall timescale. 
Using the density enhancement factor $f$, the freefall timescale and cloud mass are defined as  $t_{\rm ff} = (4\pi G f \rho_0)^{-1/2}$ and $M=fM_{\rm Jeans}$, respectively, where the Jeans mass $M_{\rm Jeans}$ is proportional to $c_{s,0}^3 \rho_0^{-1/2}$.
Thus, the mass accretion rate is proportional to $ f^{3/2}c_{s,0}^3$.
As the parameter $\alpha_0$ is proportional to $ f^{-1}$ (or $M^{-1}$), the mass accretion rate is proportional to  $ \alpha_0^{-3/2} c_{\rm s,0}^3$.
}

The model name, parameter $f$, cloud mass $M_{\rm cl}$, cloud radius $R_{\rm cl}$ and $\alpha_0$ for each model are listed in Table~\ref{table:1}. 
The parameter $C_\eta$ in the third row of Table~\ref{table:1} controls magnetic diffusivity (for details, see \S\ref{sec:eta}).
The parameters $f$ ($\alpha_0$) are varied as $f=1.4-67$ ($\alpha_0=0.01-0.5$). 
The mass accretion rate between models A ($\alpha_0=0.5$) and F ($\alpha_0=0.01$) should differ by approximately 350 ((0.5/0.01)$^{3/2}$). 
The cloud masses in the models range from $32$ to $1542\msun$.

A uniform magnetic field $B_0$ is imposed over the entire computational domain and  the initial cloud is subjected to a rigid rotation $\Omega_0$.
The initial magnetic field strength, $B_0$, and angular velocity, $\Omega_0$, are also listed in Table~\ref{table:1}, along with the ratios of rotational ($\beta_0$) and magnetic ($\gamma_0$) energies to the gravitational energy. 
The last row of Table~\ref{table:1} gives the mass-to-flux ratio $\mu$ normalised by the critical value $(2 \pi G^{1/2})^{-1}$.
In models A--F (fiducial models) and CE1--CE4 (different diffusivity models, see \S\ref{sec:eta}), the magnetic field strength $B_0$ is adjusted to have the non-dimensional parameter $\mu=2$ ($\gamma_0=0.2$).
Thus, models with different $f$ have different magnetic field strengths.
As seen from Table~\ref{table:1}, different angular velocities are used in models with different $f$ to fix the ratio of rotational to gravitational energy ($\beta_0=0.02$) among the models.
To investigate the effects of magnetic field strength on outflow properties, we develop four weak magnetic field models, CW1, CW2, EW1 and EW2 (for details, see \S\ref{sec:mag}).

\subsection{Numerical Method}
\label{sec:method}
To calculate considerably different spatial scales, we use a nested grid method in which each grid has the same number of cells (i, j, k) = (64, 64, 32) and impose mirror symmetry with respect to the $z=0$ plane (the nested grid method is detailed in \citealt{machida05a,machida05b}). 
Six grid levels ($l=1-6$) are initially prepared and the initial cloud cores are embedded in the fourth grid level ($l=4$), 
with a box size of $L_4=0.56$\,pc ($=2R_{\rm cl}$) and a cell width of $h_4=8.75\times10^{-3}$\,pc.
To avoid artificial refraction of Alf\'en waves at the computational boundary, we set a large interstellar space outside the initial cloud core. 
In each model, the interstellar media outside the initial cloud core have a uniform density of $\rho_{\rm ISM}=7.6\times10^{-21}$\,g$\cm$ ($n_{\rm ISM} = 2\times10^3\cm$) and a uniform magnetic field $B_0$. 
Note that gravity (comprising gas self-gravity and protostellar gravity) is imposed only in the region $r<R_{\rm cl}$ \citep[for details, see][]{machida10,machida11c,machida12,machida13}. 
The outer computational boundary is imposed on the surface of the first grid level $l=1$. 
The box size of the first grid level ($L_1=4.4$\,pc) is 16 times the initial cloud radius. 
The grid level is dynamically increased to maintain the Truelove condition \citep{truelove97} in which the Jeans wavelength is resolved to at least 16 grid points. 
Both the box size and cell widths are halved with each increment of the grid level. 
The maximum grid level ($l=15$) has a box size of $L_{15}=56$\,AU and a cell width of $h_{15}=0.8\,$AU.

To avoid a prohibitively short time step, we impose a sink at the centre of the computational domain. 
Protostar formation begins when the central density $n$ reaches the threshold density $n_{\rm thr}$ (set here to $n_{\rm thr}=10^{13}\cm$).   
The threshold density satisfies the necessary condition for resolving the disc forming region \citep{machida14a}. 
Once the protostar is formed, we remove the gas exceeding $n>n_{\rm thr}$ in the region  $r<r_{\rm sink}$, where $r_{\rm sink}$ is the sink (accretion) radius set to $ 1$\,AU in this study.

In models E and F, we evolve the cloud  until the protostellar mass reaches $M_{\rm ps}\sim30\msun$.
In other models, we stop the calculation when $M_{\rm ps}<30\msun$, because the accretion rate in these models is too slow to reach this mass within a reasonable time frame (typically, the calculations require from two to four months of wall-clock time).

\section{Results} 
\label{sec:results}
\subsection{Mass Accretion Rate}
\label{sec:accretionrate}
\begin{figure}
\includegraphics[width=160mm]{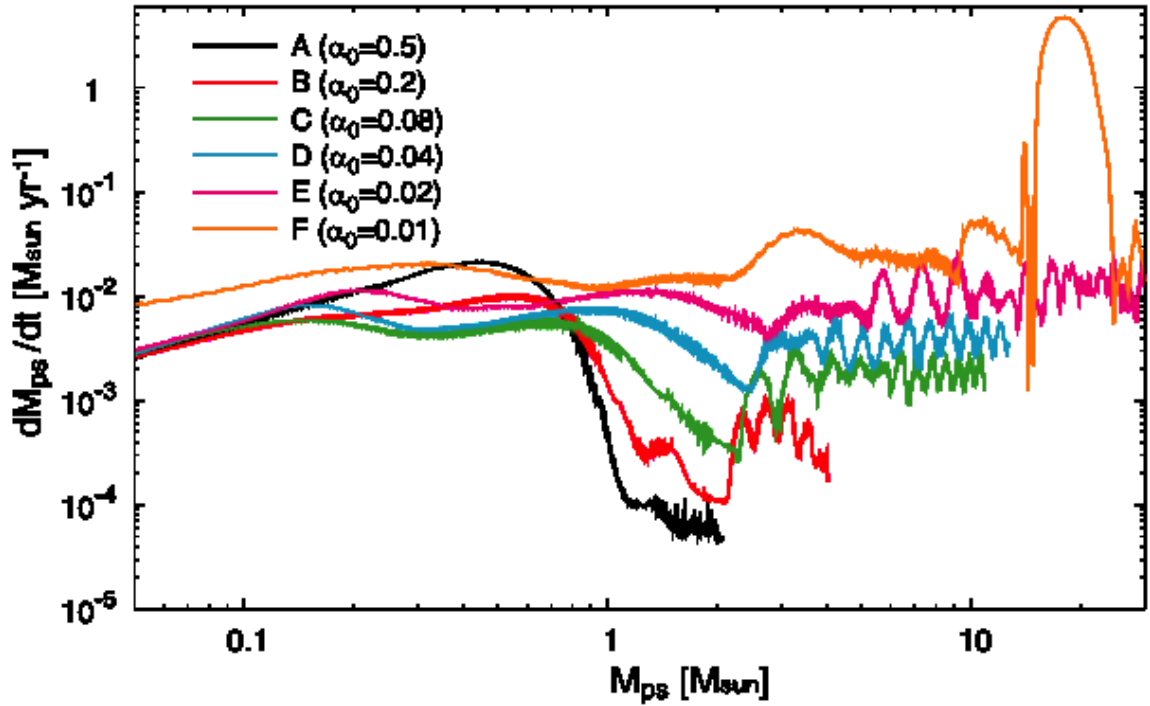}
\caption{
Mass accretion rates onto the protostar for models A--F against the protostellar mass.
}
\label{fig:1}
\end{figure}

Figure~\ref{fig:1} plots the mass accretion rate onto the protostar against the protostellar mass in models A--F. 
The figure indicates that the initially more unstable clouds have higher accretion rates. 
Within the range $M_{\rm ps}<0.1-1\msun$, the mass accretion rate is very similar in all models because it depends on the properties of the first core \citep{larson69,masunaga00}, which is supported by thermal pressure, at the very early stage. 
Just before protostar formation, the thermal pressure decreases inside the first core \citep{larson69}; just after protostar formation the central region of the first core rapidly falls onto the protostar with a higher accretion rate. 
Note that the outer region of the first core (or the first core remnant) is supported by centrifugal force and evolves into a rotationally supported disc  \citep{bate98,walch09,machida10, bate11,walch12,tsukamoto13}.

\begin{figure}
\includegraphics[width=160mm]{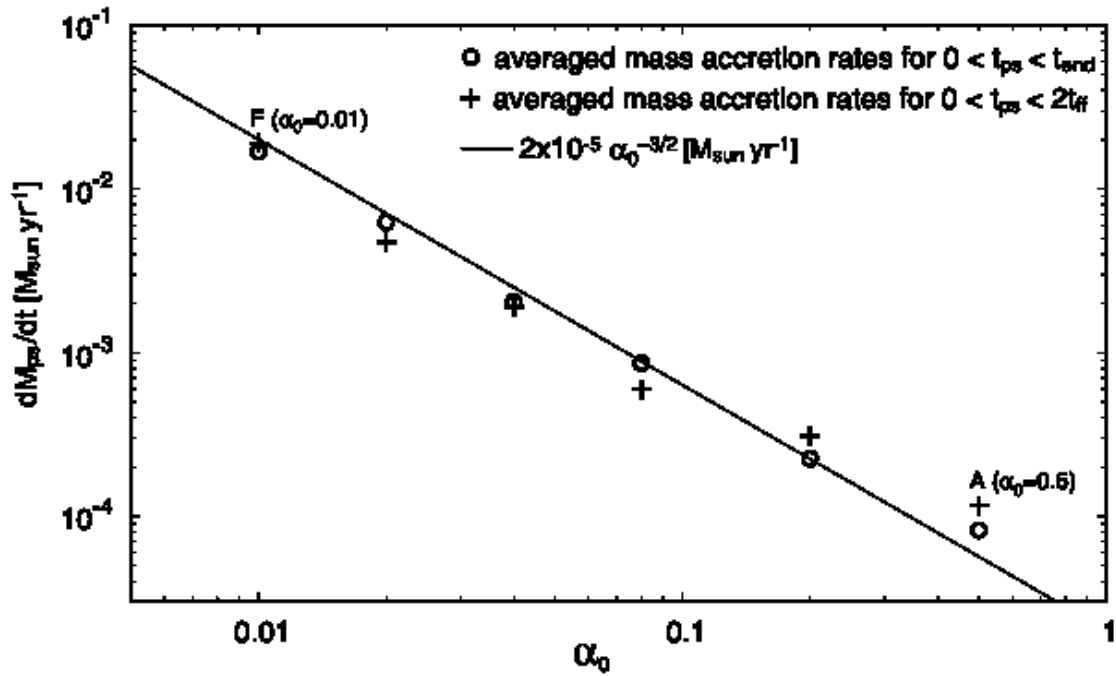}
\caption{
Mass accretion rate averaged over the time period  $ 0 < t_{\rm ps} < t_{\rm end} $ ($\bigcirc$) and $0 < t_{\rm ps} < 2\,t_{\rm ff} $ ($+$), where $t_{\rm ff}$ is the freefall timescale of each initial cloud. 
The relation $\dot{M}_{\rm ps}=2\times10^{-5}\alpha_0^{-3/2}\msun$\,yr$^{-1}$ is also plotted by the solid line. 
Model name (models A and F) and $\alpha_0$ are described in the figure.
}
\label{fig:1b}
\end{figure}

The mass accretion rate differs noticeably among the models once the protostellar mass exceeds $\sim1\msun$, with a mass accretion rate of two orders of magnitude higher in model F than in model A. 
After rotationally supported disc formation, the gas falls onto the disc and gradually moves inward towards the protostar. 
Eventually, it falls onto the protostar (or sink). 
Thus, the mass accretion rate onto the protostar is roughly determined by the mass accretion rate onto the disc, which is itself determined by the initial cloud stability (as described in \S\ref{sec:initial}). 
Figure~\ref{fig:1b} shows the averaged mass accretion rate for each model against the initial cloud stability $\alpha_0$, with the mass accretion rate averaged from just after protostar formation until the calculation ends (circle symbols) or until twice the freefall timescale elapses (cross symbols).
The figure indicates that the averaged mass accretion rate in each model corresponds closely to the theoretical prediction $\sim \alpha_0^{-3/2}\,  (c_s^3/G)$.

Figure~\ref{fig:1} also shows that the mass accretion rate drops in the range of $0.4\msun \lesssim M_{\rm ps} \lesssim 1\msun$. 
\citet{masunaga00} showed that the mass accretion rate is high in the very early main accretion phase because the gas is accreted from the inner region of the first core (remnant), where the gas temperature is high (the mass accretion rate is proportional to $\propto c_{\rm s}^3$) and is primarily supported by the thermal pressure gradient force. 
Then, after the thermally supported region of the first core is accreted, the gas accretes from the infalling envelope through the rotationally supported disc and the mass accretion rate drops. 
The oscillation of the mass accretion rate seen in Figure~\ref{fig:1} is caused by gravitational instability in the disc, as will be described later.

\subsection{Evolution of Protostellar Outflow in Typical Models}
\label{sec:typicalmodel}
\begin{figure}
\includegraphics[width=150mm]{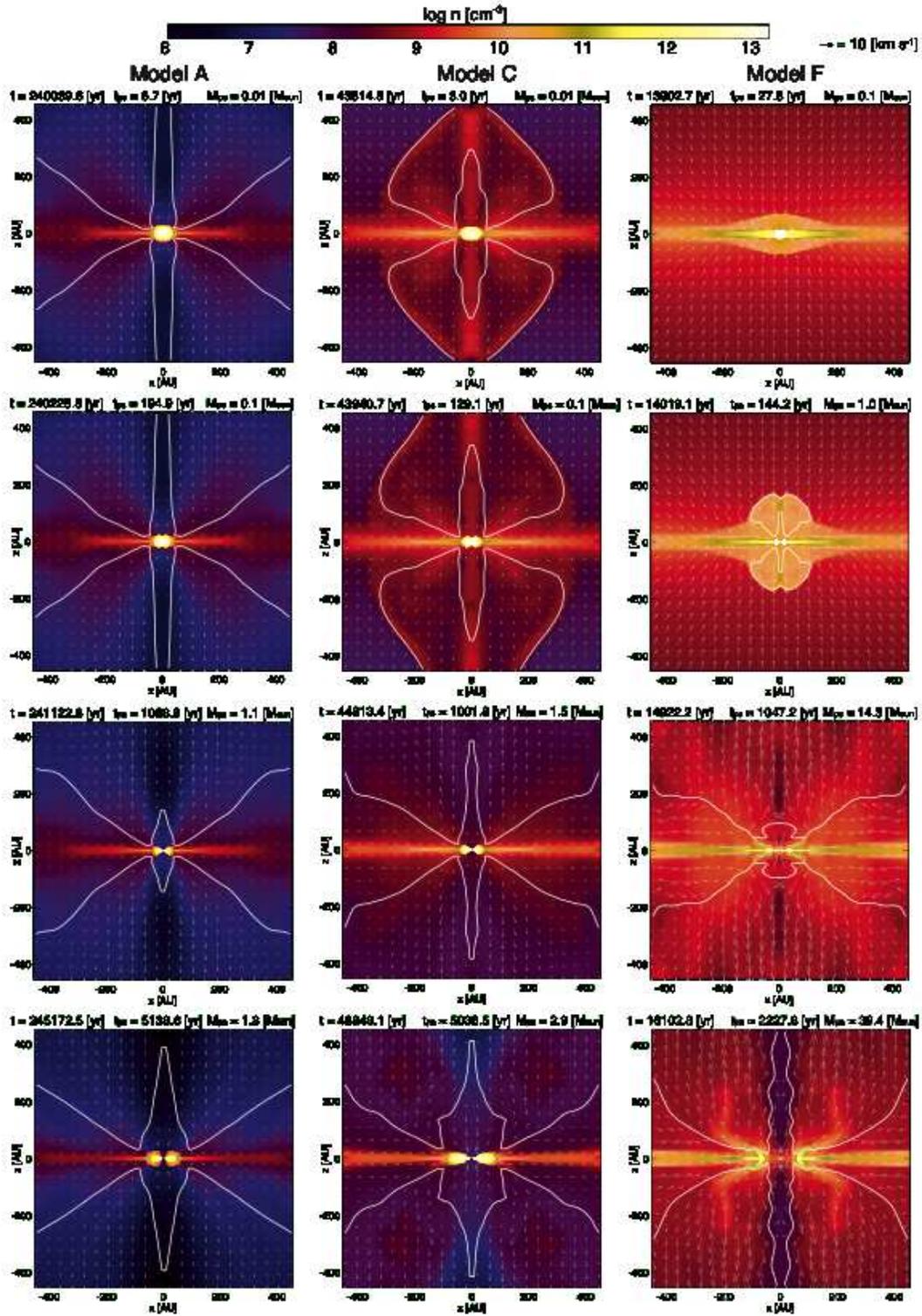}
\caption{
Time sequences of density (colour) and velocity (arrows) distributions on the $y=0$ plane in models A, C and F.
The white contours in each panel delineate the boundary between the infalling and outflowing gases, within which the gas outflows from the central region.
Above each panel is the elapsed time $t$ after the cloud begins to collapse, the elapsed time $t_{\rm ps}$ after protostar formation and the protostellar mass $M_{\rm ps}$.
The scales at the top of the figure denote density and velocity.
}
\label{fig:2}
\end{figure}

\begin{figure}
\includegraphics[width=150mm]{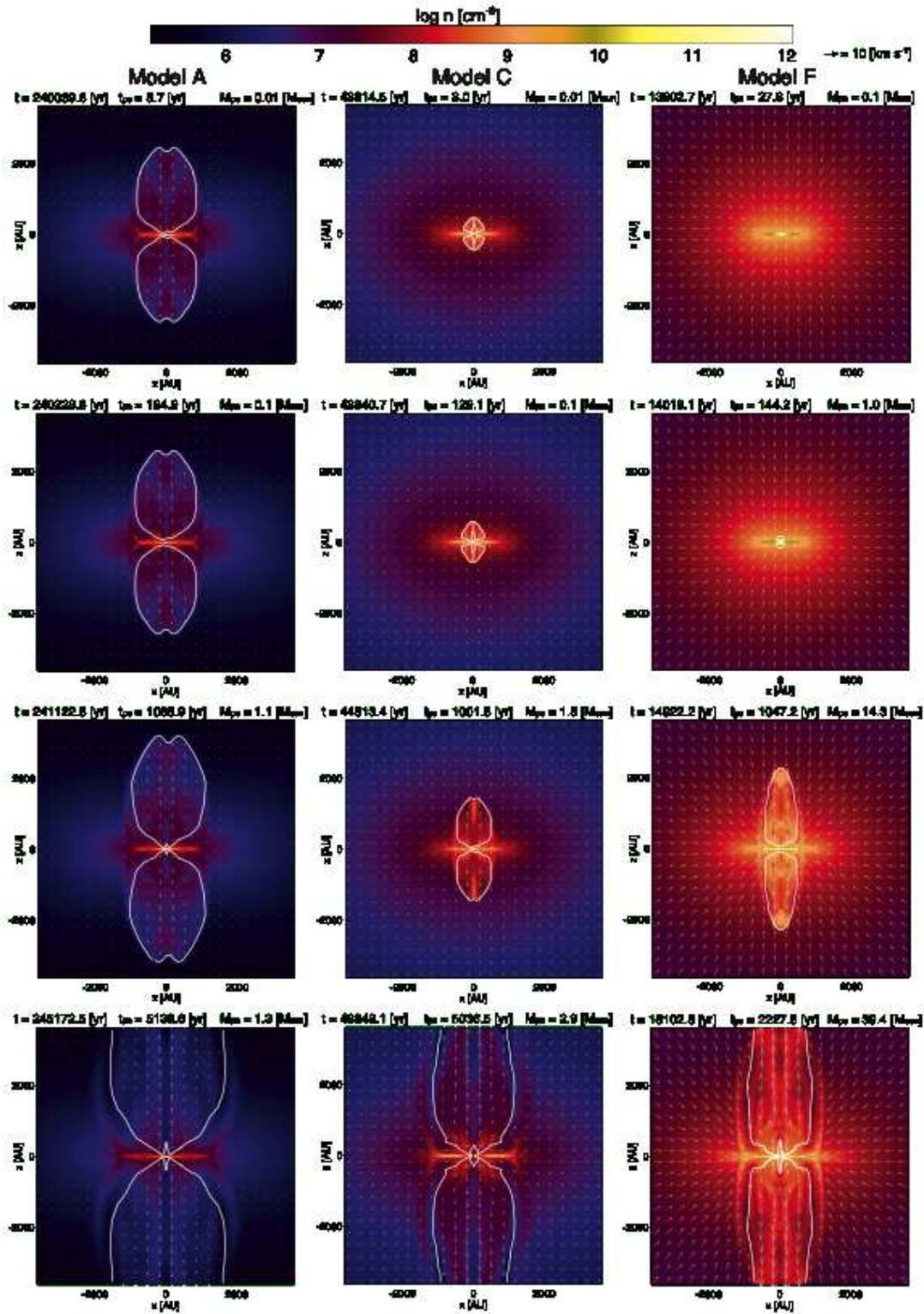}
\caption{
Fig.~\ref{fig:2} with the spatial scale enlarged by a factor of eight.
}
\label{fig:3}
\end{figure}

Figures~\ref{fig:2} and \ref{fig:3} show the time evolutions of the density and velocity distributions in models A, C and F. 
In model A (left column of both figures), the outflow already extends up to $\sim$2000\,AU at the protostellar formation epoch. 
As the accretion rate in model A is relatively low (Fig.~\ref{fig:1}), the first core sustains over a long duration. 
Note that the mass accretion rate onto the first core determines its lifetime  \citep{tomida10}. 
Initially, the wide-angle outflow is driven by the first core \citep{tomisaka02}. 
Once the protostar has formed, the outflow is driven by the  rotationally supported disc that evolves from the first core or its remnant. 
As the transition from the first core to the rotationally supported disc is smooth \citep{bate11,machida11c}, the outflow continues to be driven by the central region of the collapsing cloud after the first core formation, as seen in the left column of Figures~\ref{fig:2} and \ref{fig:3}.

At the protostar formation epoch, the outflow extends up to approximately 500\,AU in model C (central column of Figs.~\ref{fig:2} and \ref{fig:3}) but is absent in  model F (right column of Figs.~\ref{fig:2} and \ref{fig:3}). 
A higher accretion rate rapidly increases the mass (and hence the central density) of the first core, and the second collapse occurs shortly afterward. 
Therefore, a higher accretion rate shortens the lifetime of the first core \citep{saigo06,saigo08}. 
In model F, the mass accretion rate is high and the lifetime of the first core is very short; consequently, the protostar forms before the magnetic field sufficiently twists and outflow driving occurs. 
Thus, in model F the outflow evolves only after protostar formation. 
We therefore attribute the differences in the outflow evolution or size at the protostar formation epoch (first row in Figs.~\ref{fig:2} and \ref{fig:3}) to differences in the first core lifetime, which is in turn related to the mass accretion rate and initial cloud stability, $\alpha_0$.

Figures~\ref{fig:2} and \ref{fig:3} also show that the structures and shapes of the outflows are similar across the models.
In each model, the opening angle increases with time over a small scale (or disc scale) around the disc (Fig.~\ref{fig:2}), while a well-collimated outflow structure appears at the cloud core scale of  $\sim 5000$~AU (Fig.~\ref{fig:3}).
The opening angle in Figures~\ref{fig:2} and \ref{fig:3} strongly depends on both the evolutionary stage and the observed scale. 
Even when the outflow grows sufficiently (see bottom row  of Fig.~\ref{fig:3}), it is recognizable as a very wide-opening angle flow over a very small scale around the protostar (see bottom row of Fig.~\ref{fig:2}).
Note that, as the purpose of this study is to relate the accretion rate to the outflow rate (and not to investigate the evolutionary properties of outflows among models), we do not quantitatively compare the outflow shape and collimation degree among the models.

\subsection{Evolution of Circumstellar Disc in Typical Models}
\begin{figure}
\includegraphics[width=160mm]{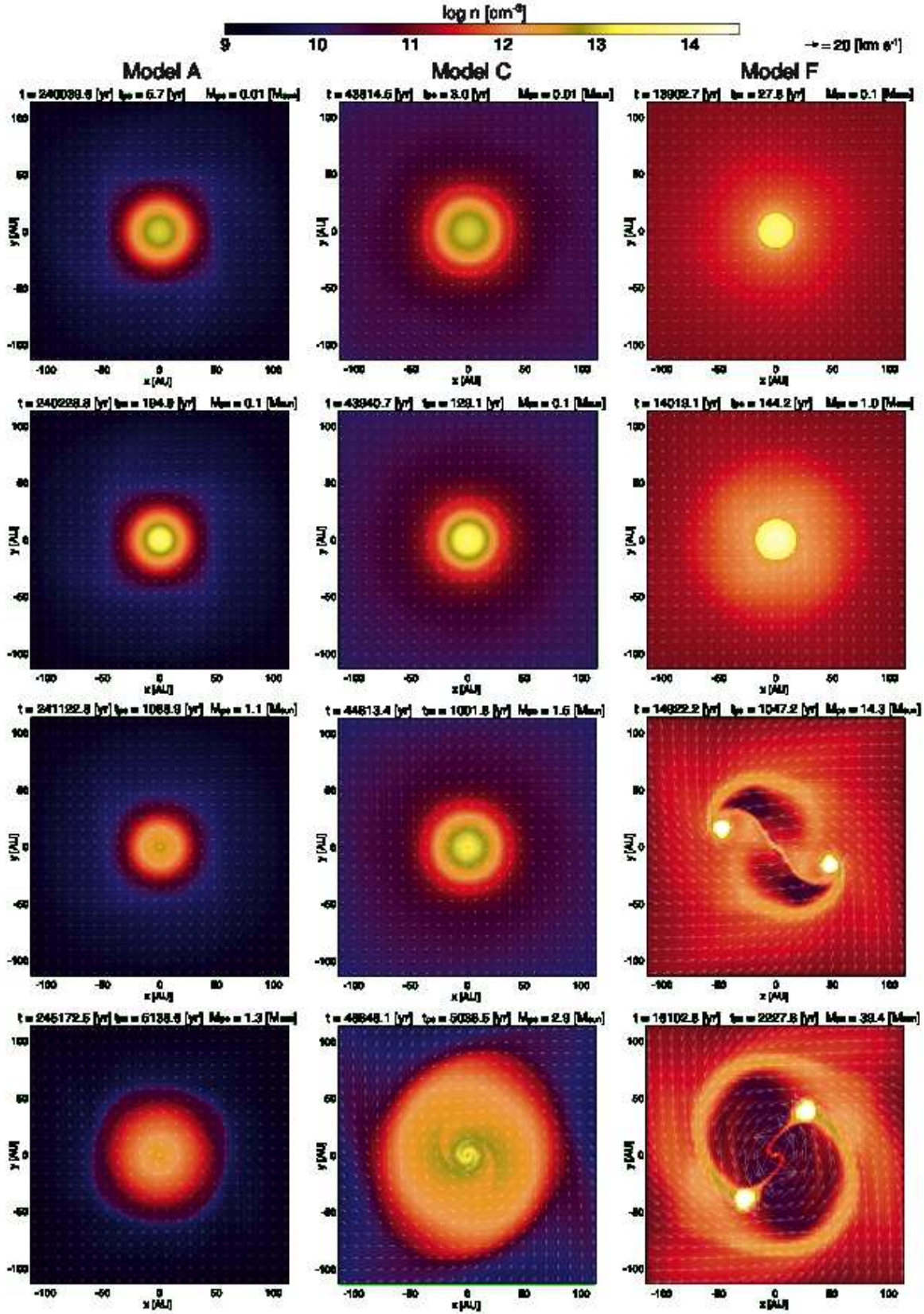}
\caption{
Time sequence of density (colour) and velocity (arrows) distributions on the $z=0$ plane in models A, C and F.
Above each panel is the elapsed time $t$  after the cloud begins to collapse, the elapsed time $t_{\rm ps}$  after protostar formation and the protostellar mass $M_{\rm ps}$.
The top scales denote density and velocity.
}
\label{fig:4}
\end{figure}

Figure~\ref{fig:4} shows the time sequence of the density and velocity distributions on the equatorial plane in models A, C and F, confirming that a higher accretion rate produces a larger disc. 
In model F, which has the largest accretion rate among the models (see Fig.~\ref{fig:1}), a large rotationally supported disc forms and then fragments.

\begin{figure}
\includegraphics[width=160mm]{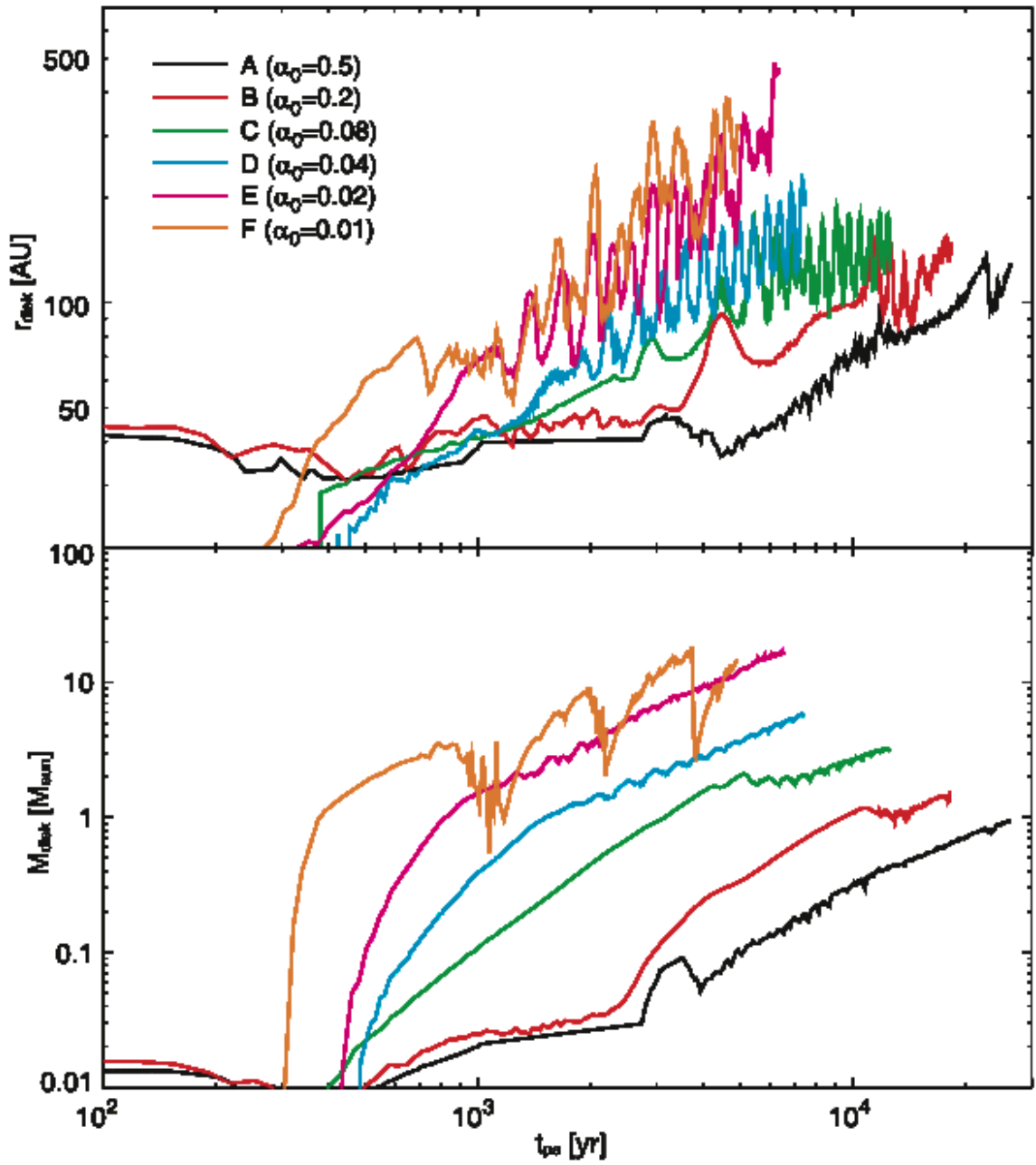}
\caption{
Radius $r_{\rm disk}$ (top panel) and mass $M_{\rm disk}$ (bottom panel) of the  rotationally supported disc in models A--F against the elapsed time after protostar formation. 
}
\label{fig:4b}
\end{figure}

To quantitatively compare the disc properties among the models, we plot the radii and masses of the  rotationally supported disks in Figure~\ref{fig:4b}. 
As a result of trial and error, we found that the disks contain all cells that fulfill the following criteria \citep{machida10,joos12}: 
\begin{enumerate}
\item the number density exceeds $n_{\rm disk}>10^9\cm$;
\item the rotational velocity is five times larger than the infall (or radial) velocity ($\vert v_{\phi}\vert >5 \vert v_{\rm r}\vert$) and larger than 90\% of the Keplerian  velocity  ($v_{\phi} > 0.9\, v_{\rm kep}$). The Keplerian velocity $v_{\rm kep}$ is defined as $v_{\rm kep}= (G M_{\rm tot}/r)^{1/2}$, where $M_{\rm tot} = M_{\rm gas} +  M_{\rm ps}$ is the sum of the high density gas  $M_{\rm gas} = \int^{n>10^{11}\cm} \rho \, dV$ and the protostellar mass $M_{\rm ps}$;
\item the rotational energy ($\rho v_{\phi}^{2}/2$) is  larger than the thermal pressure $P$ by a factor of two.
\end{enumerate} 
From Figure~\ref{fig:4b} it is seen that the models with larger accretion rates produce large massive disks in the early evolutionary stage. 
The oscillation of radius and mass seen in the figure is caused by the appearance and disappearance of spiral arms in the disc (see below). 
The rapid decrease of mass in model F is caused by the infalling of fragments to the protostar; as discussed below, fragmentation occurs only in model F.
Later, we will show that the disc size also depends on the magnetic field strength and the magnetic dissipation process (\S\ref{sec:eta}).

When the magnetic field is absent or weak, a high accretion rate causes vigorous fragmentation \citep[e.g.][]{hennebelle11} because the disc surface density increases within a short time, while the Toomre's $Q$ value ($\equiv c_s \kappa/(\pi G \Sigma)$,\citealt{toomre64}) decreases. 
However, under a strong magnetic field, outflow and magnetic braking will effectively transport the angular momentum outward, promoting mass accretion from the disc to the protostar \citep{seifried11,joos12,joos13,machida14a}. 
Therefore, the disc surface density decreases, the $Q$ value increases, and the disc consequently stabilises. 
In the massive star formation process, vigorous fragmentation within the disc (or over a small scale) should be suppressed because each fragment would be expected to evolve into a low- or intermediate-mass star without forming a (very) massive star  \citep[see also][]{hennebelle11,peters11,myers13}. 
Note that fragments tend to fall onto more massive protostars; therefore, a few massive protostars may continue to acquire mass and evolve into very massive stars  \citep{machida11a,vorobyov15,machida15,sakurai16}. 
With an initially strong magnetic field (models A--F), fragmentation occurs when the mass accretion rate is as high as $\dot{M} \gtrsim 10^{-2}\msun$\,yr$^{-1}$ (model F), while fragmentation is suppressed with an accretion rate of $\dot{M} \lesssim 10^{-2}\msun$\,yr$^{-1}$(models A--E).
However, there are few fragmentation events in model F, and therefore this system should evolve into a binary or a few-star multiple system rather than a small stellar cluster comprising several tens of stars. 
Note that, as the fragmentation condition and Toomre's $Q$ parameter depend on the gas temperature, we should also carefully consider fragmentation with thermodynamics in addition to the mass accretion rate or disc surface density (see also \S\ref{sec:feedback}).

\begin{figure}
\includegraphics[width=160mm]{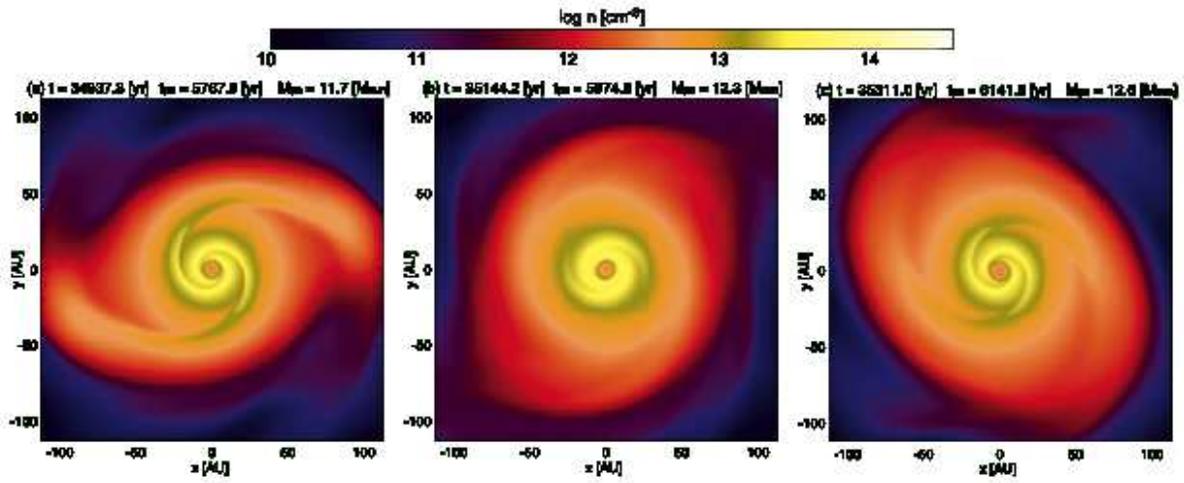}
\caption{
Time sequence of density (colour) distribution on the $z=0$ plane in model D.
Above each panel is the elapsed time $t$ after the cloud begins to collapse, the elapsed time $t_{\rm ps}$ after protostar formation  and the protostellar mass $M_{\rm ps}$.
The top scale denotes density.
}
\label{fig:5}
\end{figure}

Although no fragmentation occurs in models A--E, a higher accretion rate induces gravitational instability and establishes a spiral structure in the disc. 
Figure~\ref{fig:5} shows the time sequence of density distributions on the equatorial plane in model D. 
Once the spiral structure develops (Fig.~\ref{fig:5}{\it a}), the gas in the disc rapidly falls onto the central protostar under the effective angular momentum transfer via the gravitational and pressure torques. 
The disc transiently stabilises against the self-gravity (Fig.~\ref{fig:5}{\it b}), until the high mass accretion rate onto the disc increases the disc surface density, and the spiral structure reforms (Fig.~\ref{fig:5}{\it c}). 
The gravitational instability of the disc explains the non-steady accretion seen in Figure~\ref{fig:1}. 
This phenomenon has been thoroughly investigated in low-mass star formation processes \citep[e.g.][]{vorobyov06,tomida17}. 
Thus, both magnetic effects and gravitational instability contribute to the angular momentum transfer.

\subsection{Mass Inflow and Outflow Rates}
\label{sec:morates}
\begin{figure}
\begin{center}
\includegraphics[width=160mm]{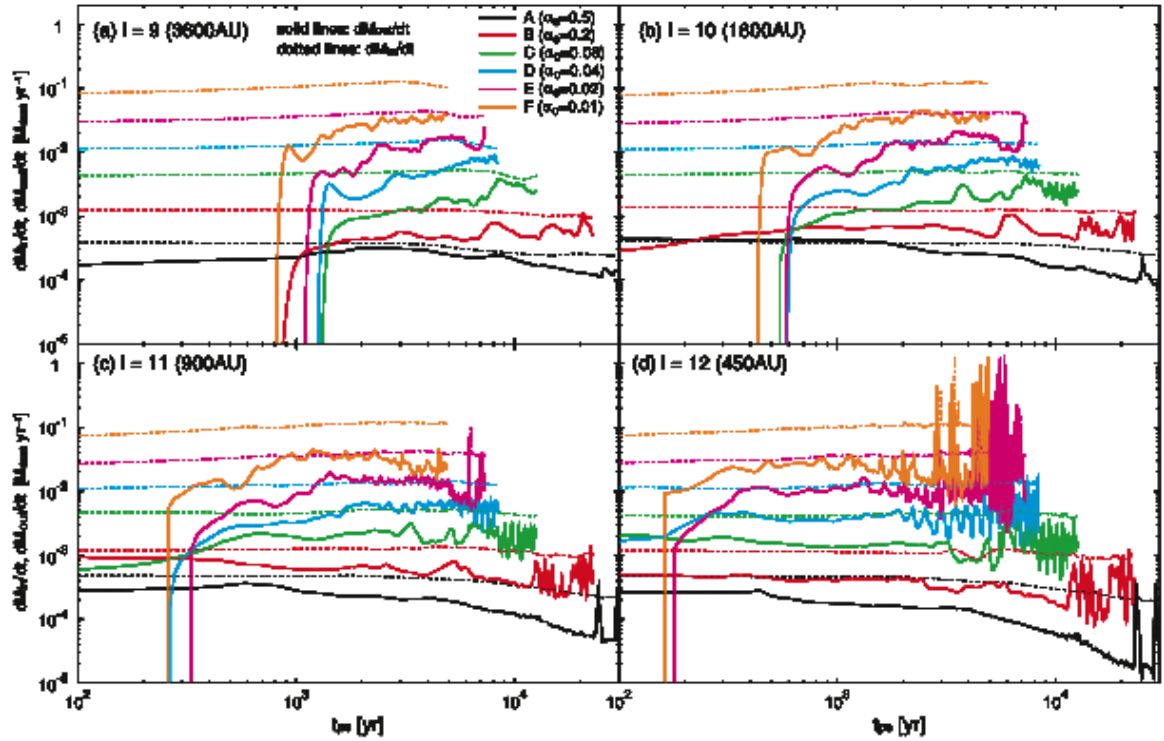}
\caption{
Mass inflow (dotted lines) and outflow (solid lines) rates in models A--F at different scales (panels {\it a}-{\it d}) versus elapsed time after protostar formation. Both rates are estimated on the surface of each nested grid.
The grid level and box size are described in each panel.
}
\label{fig:6}
\end{center}
\end{figure}

To investigate the efficiency of mass ejection, we plot the mass inflow and outflow rates in Figure~\ref{fig:6}. 
These rates are estimated on the cell surfaces of different grids (for details, see \citealt{tomisaka02}). 
On the $l=9$ grid (Fig.~\ref{fig:6}{\it a}), the inflows and outflows cover 1/32 of the initial cloud (3500\,AU), whereas on the level $l=12$ grid (Fig.~\ref{fig:6}{\it d}) they cover the entire disc region (450\,AU). 
Clearly, the infall rate onto each grid depends on the cloud parameter $\alpha_0$ and strongly affects the mass outflow rate (the higher the accretion rate, the higher the outflow rate). 
Figure~\ref{fig:6} also shows that the respective infall rate of the models do not significantly depend on the scale of the collapsing region (compare the infall rate of each model in panels ({\it a})-({\it d})).
Moreover, the outflow rate is time-variable in all models and is exaggerated on small scales (or disc scales). 
This indicates that the outflow is intermittently driven by the circumstellar disc.
As shown in Figure~\ref{fig:5}, mass accretion on the protostar is promoted by a non-axisymmetric structure that develops in the disc. 
Thus, the disc time variability caused by gravitational instability also significantly affects outflow driving. 
On the disc scale (Fig.~\ref{fig:6}{\it d}), the outflow time variability is significant in models with higher accretion rates (models E and F). 
The high accretion rate gravitationally destabilises the disc over short time intervals, generating a highly time-variable outflow.

\begin{figure}
\begin{center}
\includegraphics[width=160mm]{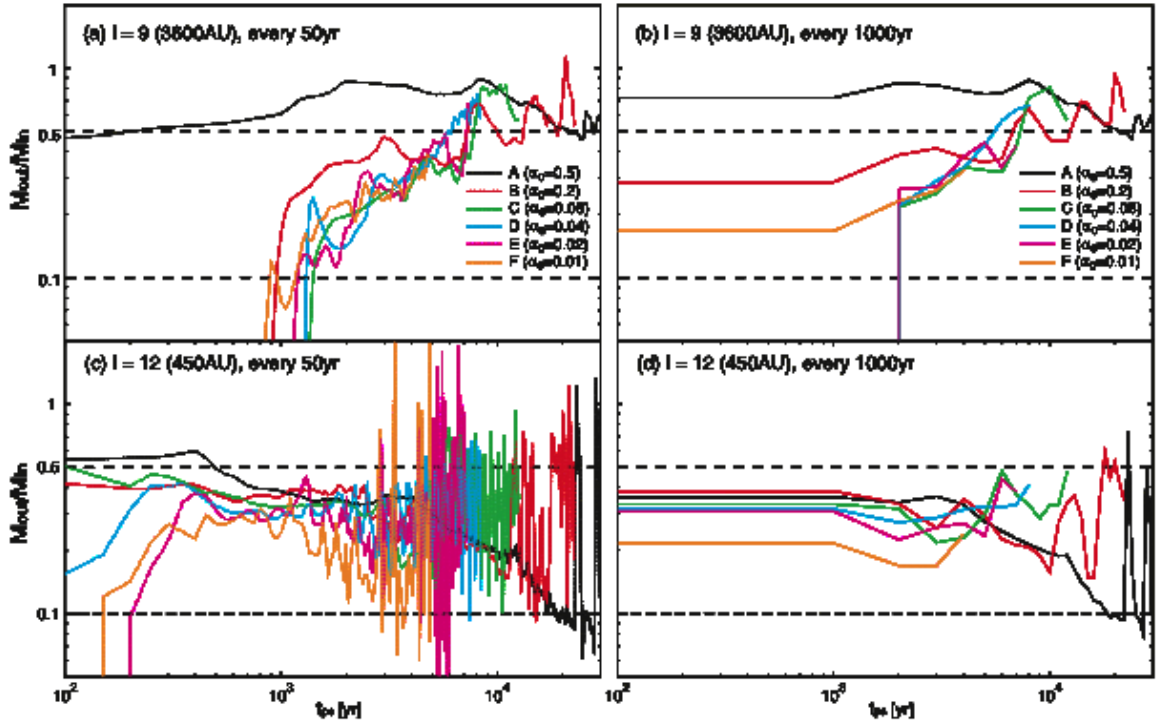}
\caption{
The ratio of outflow to inflow masses plotted at scales of 3600 AU (top panels) and 450 AU (bottom panels). 
The ratio is averaged over each 50 (left panels) and 1000\,yr (right panels) interval.
Lower and upper black dashed lines indicate $M_{\rm out}/M_{\rm in}=0.1$ and 0.5, respectively.
}
\label{fig:7}
\end{center}
\end{figure}

Figure~\ref{fig:7} plots the ratios of outflow to inflow masses, averaged every 50 (left panels) and 1000\,yr (right panels). 
To derive the ratio $M_{\rm out}/M_{\rm in}$, the inflow $M_{\rm in}$ and outflow $M_{\rm out}$ masses of each grid are, respectively, estimated as
\begin{eqnarray}
M_{\rm out} = \int^{t' + \Delta t}_{t'} \frac{dM_{\rm out}}{dt} dt, \\
M_{\rm in} = \int^{t'+\Delta t}_{t'} \frac{dM_{\rm in}}{dt} dt, 
\end{eqnarray}
where $dM_{\rm out}/dt$ and $dM_{\rm in}/dt$ are the mass inflow and outflow rates of each grid (Fig.~\ref{fig:6}) and  $\Delta t=$50\,yr (left panels) or 1000\,yr (right panels).
At first, the protostar formation epoch $t_{\rm ps} = 0$ is set to $t'$ and the mass inflow and outflow rates are integrated over $\Delta t$.
Then, $t' = t'_{\rm pre}+\Delta t$ is used as the next integration and the integration is repeated up to the end of the simulation data, where $t'_{\rm pre}$ is the previous lower limit of integration.
The outflow is driven near the disc. 
On the disc scale (Figs.~\ref{fig:7}{\it c} and {\it d}), the ratios range from approximately 0.2 to 0.4, indicating that about 20-40\% of the infalling gas is ejected from the disc as outflow. 
The ejection rate estimated from our study corresponds closely to previously reported rates \citep{tomisaka98,tomisaka02,hennebelle08,seifried12}. 
In the large scale, the ratio of outflow to inflow mass increases with time because the outflow opening angle at small scales widens with time, sweeping a larger fraction of the infalling gas (see Figs.~\ref{fig:2} and \ref{fig:3}). 
The ratios reach or exceed 50\% (Figs. ~\ref{fig:7}{\it a} and {\it b}) on the large scales (3600 AU), implying that the protostellar outflow finally ejects over half of the infalling gas from the star-forming cloud into interstellar space.

\subsection{Comparison with Observations}
\label{sec:comp-obs}
The process of protostellar outflows in low-mass star formation has been well investigated in many observational and theoretical studies and is known to be driven by magnetic effects \citep[e.g.][]{konigl00,pudritz07,arce07,frank14}.
Although there have been only a limited number of studies focusing on massive outflow in the massive star formation process, these are also considered to be driven by magnetic effects \citep{hennebelle11,seifried12}.
Recent observations also imply the importance of the magnetic field and protostellar outflows in massive star forming regions \citep{tan13,li15,tan16}. 
Therefore, to better clarify the massive star formation process, we need to further investigate the effects of the magnetic field on massive outflows.
This study focuses on the relations between the outflow and accretion rates and the outflow properties, with results that can hopefully provide insights into massive star formation. 
For this reason, we compare our results with observed massive outflows in massive star forming regions.

\begin{figure}
\includegraphics[width=160mm]{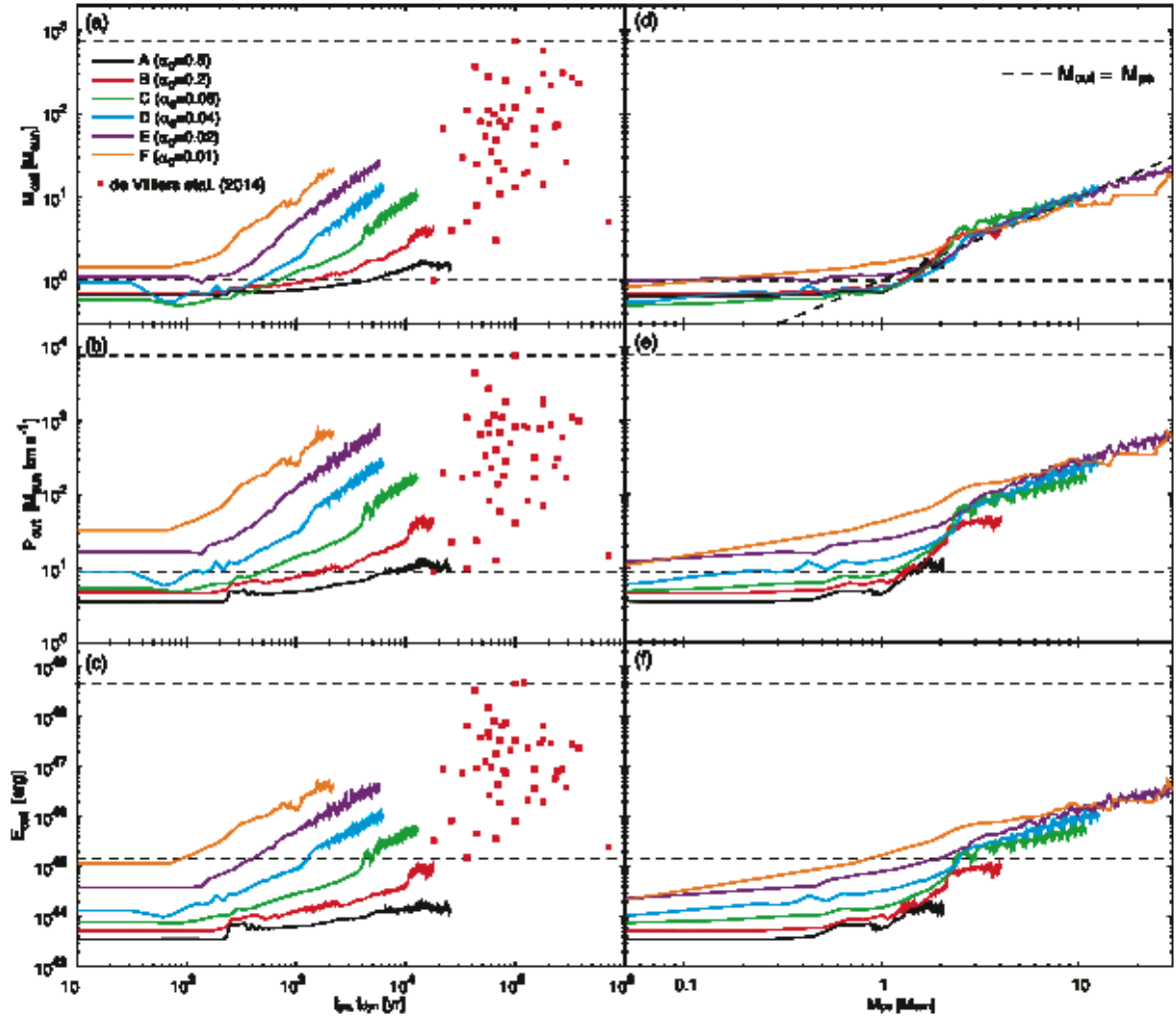}
\caption{Outflow mass ({\it a}, {\it d}), outflow momentum ({\it b}, {\it e}) and outflow kinematic energy ({\it c}, {\it f}) in models A--F (solid lines) against the elapsed time after protostar formation [left panels, ({\it a})-({\it c})] and protostellar mass  [right panels, ({\it d})--({\it f})]. 
Red dots in the left panels are observational data taken from Table 5 of \citet{villiers14}. 
Horizontal dashed lines mark the maximum and minimum of the observational data.
The relation $M_{\rm out}= M_{\rm ps}$ is plotted in panel ({d}).
}
\label{fig:8}
\end{figure}

Figure~\ref{fig:8} shows the outflow mass $M_{\rm out}$, momentum $P_{\rm out}$ and kinetic energy $E_{\rm out}$ as functions of time (left panels) and protostellar mass (right panels). The results are presented as solid lines. 
In our simulations, these quantities are, respectively, estimated as
\begin{equation}
M_{\rm out} = \int_{v_r > v_{\rm cri}} \, \rho \,  dV,
\end{equation}
\begin{equation}
P_{\rm out} = \int_{v_r > v_{\rm cri}} \,  \rho\,\vert \vect{v} \vert \, dV,
\end{equation}
\begin{equation}
E_{\rm out} = \int_{v_r > v_{\rm cri}} \,  \rho\, \vect{v}\cdot\vect{v} \, dV,
\end{equation}
where we integrate each quantity at each grid point (over the entire computational domain) with radial velocity $v_r > v_{\rm cri}$. 
Through trial-and-error, we determined $v_{\rm cri}$ to be $0.5$\,km\,s$^{-1}$. 
In this determination, we estimated the outflow physical quantities only without integrating other components such as the circumstellar disc.
Note that, by setting $v_{\rm cri}= 0.5$\,km\,s$^{-1}$, we somewhat underestimate the outflow physical quantities because we ignore the low-velocity outflow components ($0<v_{\rm out}<v_{\rm cri}$). 
As a reference, the left panels of Figure~\ref{fig:8} plot the most recent observational data taken from \citet{villiers14}. 
These observations are plotted against the outflow dynamical timescale $t_{\rm dyn}$ (defined as the outflow length $L_{\rm out}$ divided by the typical outflow velocity $v_{\rm out,typ}$, namely $t_{\rm dyn} = L_{\rm out}/v_{\rm out,typ}$). 
We find that the simulated outflows and their maximum ($v_{\rm out, max}$) and typical ($v_{\rm typ}$) velocities strongly vary with time.
Thus, we cannot easily understand the time evolution of the outflow physical properties in light of the dynamical timescale of the simulation because the dynamical timescale ($t_{\rm dyn} = L_{\rm out}/t_{\rm out,max}$ or $L_{\rm out}/t_{\rm out,typ}$) is not a simply increasing function of time.  
For this reason, the definition of time (the abscissa axis in the left panels of Fig.~\ref{fig:8}) differs between our calculations (elapsed time after the protostar formation) and observations (outflow dynamical timescale), and further effort may be therefore be needed to make a valid comparison.

As shown in Figure~\ref{fig:8}, the simulated quantities gradually increase as functions of both time (left panels) and protostellar mass (right panels). 
Models with higher accretion rates or smaller $\alpha_0$ increase more rapidly over time (Fig.~\ref{fig:8}, left panels). 
This result can be explained in the context of the ratio of the outflow to the inflow rate. 
As shown in Figure~\ref{fig:7}, this ratio is almost identical among the models and, therefore, increasing the accretion rate naturally generates a massive outflow with high momentum and kinetic energy.
Because these simulated outflow quantities continue to increase, they will reach the observed values of \citet{villiers14} at a further evolutionary stage.

On the other hand, each quantity seems to converge to a certain value in the right panels of Figure~\ref{fig:8}, in which the simulated quantities are plotted against the protostellar mass.
The protostellar mass roughly corresponds to the mass accretion rate (or inflow rate) multiplied by the elapsed time after protostar formation ($M_{\rm ps} \sim \dot{M} \times t_{\rm ps}$). 
The convergence in these physical quantities means that the outflow properties are primarily determined by the mass falling onto the central region, which is an expected result as the outflow is powered by the release of gravitational energy, which is proportional to $\dot{M} \times t_{\rm ps}$.

\subsubsection{Outflow Mass}
We first compare the respective outflow masses produced by the simulations and observations. 
In a statistical study of outflow observations, \citet{wu05} showed that the outflow mass is widely distributed over the range $10^{-3}\msun \lesssim M_{\rm out} \lesssim 10^3\msun$, which includes both low- and high-mass stars. 
\citet{beuther02} observed 26 high-mass star forming regions and identified 21 outflows driven from protostars, with masses ranging from $0.04$ to $1423\msun$. 
Nearly all of these outflows massed from $\sim1$ to $100\msun$  (average $101\msun$), with the wider range attributable to a few extremely massive and extremely low-mass outflows.
\citet{zhang05} identified 35 outflows among 69 luminous $IRAS$ sources. 
They showed that the masses of massive protostellar outflows were tens of $\msun$, with an average of $M_{\rm out, ave}=20.6\msun$. 
Recently, \citet{maud15} observed 99 massive young stellar objects and confirmed massive outflows with a mass range of $\sim10-100\msun$.
\citet{villiers15} listed 44 molecular outflows appearing in $^{13}$CO lines and 6.7-GHz methanol masers.
They plotted the outflow mass distributions in their own and previous studies \citep{beuther02,wu04,zhang05} and reported massive outflows with typical masses of $ \sim 10-100 \msun$ and an overall mass range of $10\msun \lesssim M_{\rm out} \lesssim 10^3\msun$ in massive protostars. 
The majority of their investigated outflows also massed $\sim100\msun$. 
Note that, as massive outflows are easily observed, they may have been preferentially found in these observations.

As already described, a massive protostar has a massive outflow.
This tendency does not contradict the observed star formation efficiency of $\sim30-50$\% \citep[e.g.][]{andre10}. 
Observations have also shown that the outflow mass increases with protostellar luminosity or protostellar mass \citep{wu04,villiers14,maud15}. 
In the main accretion phase, protostellar luminosity is proportional to $\propto M_{\rm ps} \dot{M}$. Thus, the observations imply that massive outflows arise from massive protostars or high mass accretion rates. 
Although we terminate the calculation when the protostellar mass reaches $\sim30\msun$ in models with higher mass accretion rates, the simulated outflow masses ($\sim20-30\,\msun$) agree reasonably with the observations ($\sim10-100\msun$).

\subsubsection{Outflow Momentum and Energy} 
To more precisely compare the simulated and observed results, we plot the momenta and energies of the outflows derived by simulations in Figures~\ref{fig:8}{\it b}, {\it e}, {\it c} and {\it f}. 
Figures~\ref{fig:8}{\it b} and {\it c} also show the observed outflow momenta and energies from \citet{villiers14}. 
In their statistical study of both low- and high-mass  protostars, \citet{wu05} showed that the momenta of observed outflows range from $10^{-3}$ to $10^4 \msun\, {\rm km}\, {\rm s}^{-1}$.  
In our simulation, the outflow momentum depends on the time and on $\alpha_0$, but the massive protostars have outflow momenta of $3 \lesssim P_{\rm out}/(\msun\, {\rm km}\, {\rm s}^{-1}) \lesssim 1000$ (Fig.~\ref{fig:8}{\it b} and {\it e}). 
Our derived outflow momenta and kinetic energies closely agree with other observations of massive outflows \citep{beuther02,zhang05,maud15}; however, they are somewhat smaller than the observed values. 
Nevertheless, the simulated outflows continue to be driven, indicating that the outflow momentum and {kinetic energy} 
would increase at a further evolutionary stage.
Similarly, our derived massive outflow kinetic energies approach those in the observations (Fig.~\ref{fig:8}{\it c}) and would be expected to converge to the observed values under continuation of the simulations. 
In addition, the outflow kinetic energies derived from our simulations are comparable to observations by \citet{beuther02}, \citet{zhang05} and \citet{maud15}.
Note that, to estimate the outflow kinetic energy in the late accretion stage, we need to consider the effect of 
 radiative feedback.
Note also that observational studies will preferentially detect more massive (or more evolved) outflows. 
Thus, for a proper comparison of simulated and observed massive outflows, we require longer-term simulations or more spatially resolved observations (which would identify younger outflows).

\subsubsection{Outflow Momentum Flux} 
\begin{figure}
\includegraphics[width=160mm]{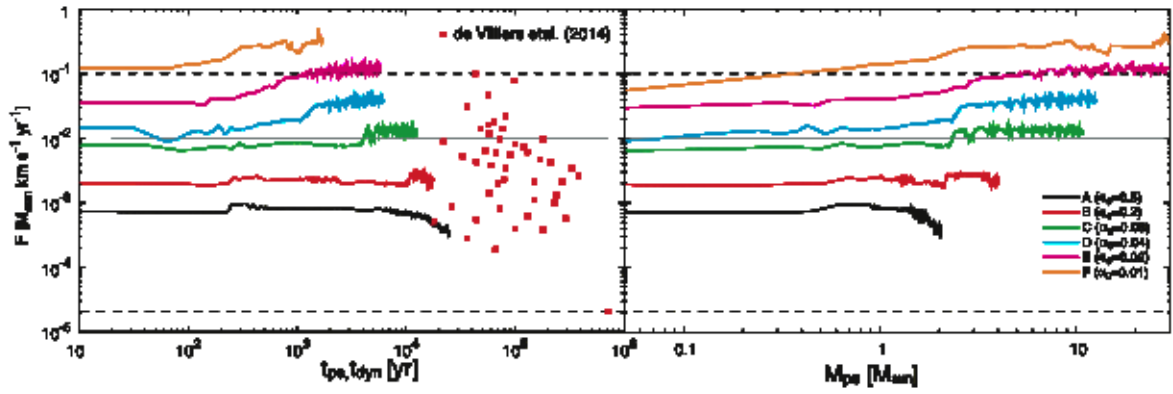}
\caption{
Outflow momentum flux in models A--F against elapsed time after protostar formation (left) and protostellar mass (right). 
Red dots plotted in the left panel are observational data taken from Table 5 of \citet{villiers14}. 
Horizontal dashed lines mark the maximum and minimum of the observational data. 
}
\label{fig:9}
\end{figure}

The outflow momentum flux $F$ can be used to understand the driving mechanism or engine of the outflow \citep{bally83,cabrit90,cabrit92,bontemps96}.
Observations have determined the outflow momentum flux as $F_{\rm out}=P/t_{\rm dyn}$ (note that the outflow momentum flux has the dimensions of force). 
The momentum flux in each of our simulated models, derived as 
\begin{equation}
F = P_{\rm out}/t_{\rm out},
\end{equation}
is plotted in Figure~\ref{fig:9}. 
Here, $t_{\rm out}$  is the elapsed time after the outflow emerges. 
The left panel of Figure~\ref{fig:9} also plots the momentum fluxes listed in Table~5 of \citet{villiers14}.

After the outflow emergence, the outflow momentum flux in each model remains nearly constant, with a slight increase when $M_{\rm ps}\gtrsim 1\msun$ except in model A. 
Clearly,  the momentum flux is larger in models with higher accretion rates (or smaller $\alpha_0$) than in models with lower accretion rates (or larger $\alpha_0$), which is to be expected because the gravitational energy is converted into magnetic and rotational (or kinetic) energies and the outflow is driven by the Lorentz and centrifugal forces \citep{blandford82,uchida85}. 
Thus, as the accretion rate increases, a strong magnetic field and rapid rotation develop around the protostar or circumstellar disc, yielding a powerful outflow.

The simulated momentum fluxes range from $10^{-4}$ to $0.3 \msun \, {\rm km}\, {\rm s}^{-1}\, {\rm yr}^{-1}$.  
The simulation results are consistent with the observations of \citet{villiers14} (Fig.~\ref{fig:9}, left panel) and with those of \citet{beuther02}, \citet{zhang05} and \citet{maud15}. 
From these agreements, the driving mechanism of massive outflow may be explained as analogous to the magnetically driven outflow in the low-mass star formation process.
Because the momentum flux does not (significantly) depend on the evolutionary stage, we may estimate the mass accretion rate from the observed momentum flux.

\section{Discussion}
\subsection{Effects of Magnetic Dissipation on Outflow Driving}
\label{sec:eta}
Whereas the protostellar outflow is considered to be driven by magnetic effects (magnetocentrifugal or magnetic pressure-driven winds), the magnetic field dissipates in the high-density gas region in which the ionisation degree is very low \citep[e.g.][]{nakano80,nakano86,nakano95,nakano02,tomida13,tsukamoto15a,tsukamoto15a,tomida15}. 
Thus, to investigate the outflow driving process, we must carefully consider the magnetic dissipation. 
Here, we include ohmic dissipation and adopt the dissipation coefficient formulated in \citet{machida07} and \citet{machida08}. 
The coefficient $\eta_{\rm fid}$ in equation~(\ref{eq:etadef}) is essentially determined by the degree of gas ionisation, which significantly depends on the dust properties. 
For instance, the ionisation degree increases and the dissipation coefficient decreases with dust growth because the dust surface area decreases and charged particles do not efficiently stick to the dust grains. 
Thus, to correctly determine the magnetic dissipation coefficient, it is necessary to solve the dust growth and chemical reactions of charged particles in the MHD calculations. 
However, present computational resources are insufficient for this task.

To investigate the effect of magnetic dissipation on outflow driving, we vary magnetic diffusivity as
$\eta = C_\eta\, \eta_{\rm fid}$, ranging parameter $C_\eta$ from 0.01 to 100 (models C, CE1, CE2, CE3, CE4) (see Table~\ref{table:1} and \S\ref{sec:basic}). 
Among the models, $C_{\rm \eta}$ varies by a factor of $10^4$. 
Because the magnetic Reynolds number $Re_{\rm m}$ is roughly proportional to $\rho^{-1}$ (see Fig. 1 and the corresponding description in \citealt{machida07}), the difference in the density of the magnetic dissipation region between models CE1 ($C_{\eta}=0.01$) and CE4 ($C_{\eta}=100$) is about $10^4$. Simply assuming constant temperature and the Jeans wavelength as the typical spatial scale, the spatial scales of the magnetic dissipation regions differ by approximately $10^2$. 
Note that in reality, the temperature varies and estimating the differences in the magnetic dissipation regions is non-trivial \citep{machida07,dapp10}.

\begin{figure}
\includegraphics[width=160mm]{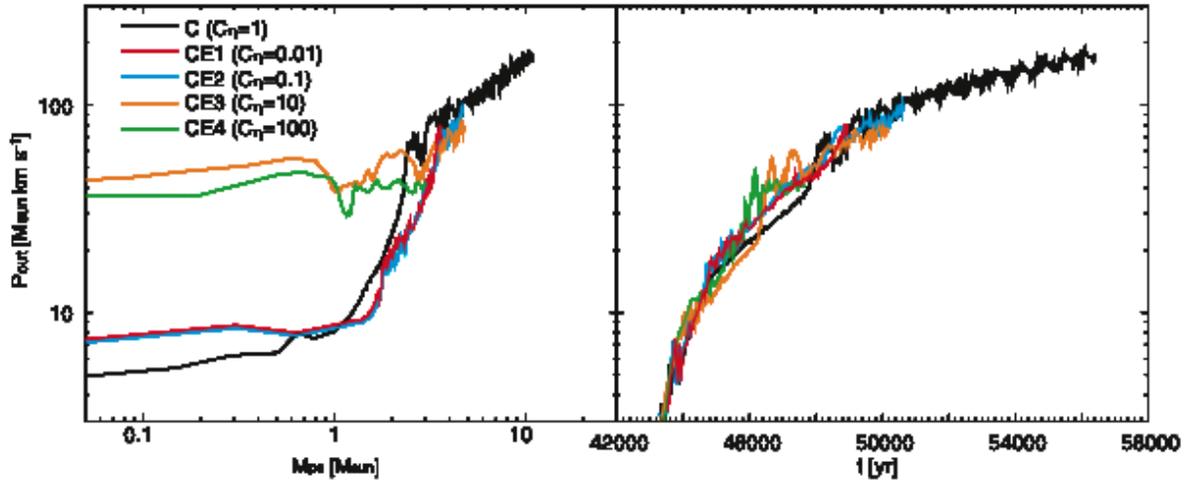}
\caption{
Outflow momenta in models with $C_{\eta}$ = 0.01 (CE1), 0.1 (CE2)  ,1 (C), 10 (CE3) and 100 (CE4) versus protostellar mass (left) and elapsed time after the cloud begins to collapse (right).
}
\label{fig:10}
\end{figure}

We then calculate the cloud evolution by fixing the parameters to those of model C but varying the diffusivity $\eta$ (i.e. $C_\eta$). 
The outflow momentum for each model is plotted against the protostellar mass in the left panel of Figure~\ref{fig:10}.
In this panel, the outflow momenta are larger in models CE3 and CE4 than in models C, CE1 and CE2 for $M_{\rm ps}\lesssim 3\msun$. 
As models CE3 and CE4 have larger diffusivities, the magnetic field of the first core, which forms before protostar formation, is weak.
Thus, in these models, angular momentum is not effectively transported by magnetic effects and the first core sustains over a long duration with high rotational energy \citep{saigo06,saigo08}. 
The outflow, which appears just after the first core formation \citep{machida04,tomisaka02,tomida13,tomida15}, travels a large distance during the long lifetime of the first core and possesses a large momentum prior to protostar formation. 
The right panel of Figure~\ref{fig:10} plots the outflow momentum  against the elapsed time following the beginning of the cloud collapse. 
As the outflow momenta are very similar among the models, we can attribute the differences in the left panel of Figure~\ref{fig:10} to the first core lifetimes. 
In summary, the diffusivity affects the first core lifetime, but does not significantly influence the outflow momentum.

\begin{figure}
\includegraphics[width=160mm]{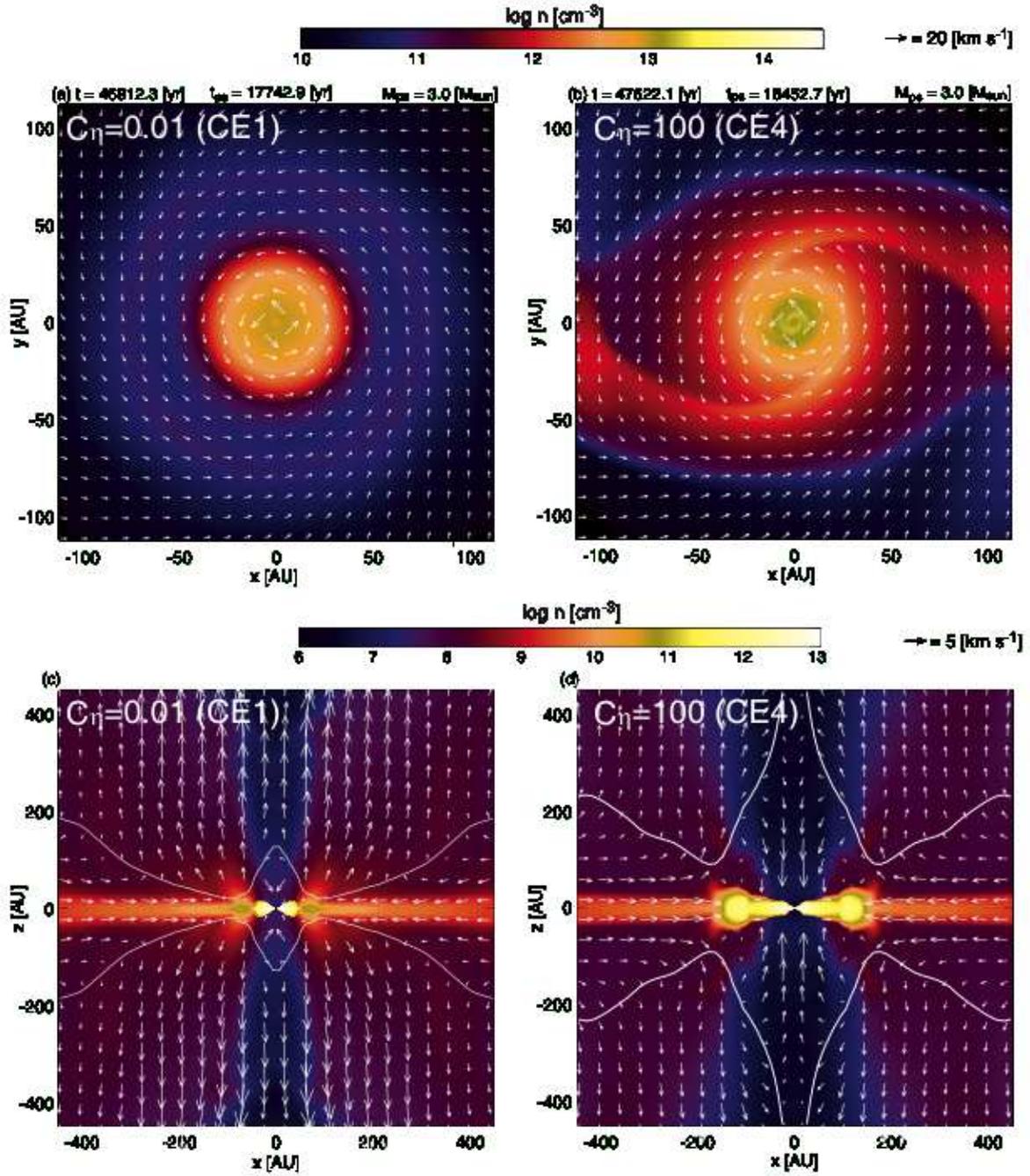}
\caption{
Density and velocity distributions on the equatorial plane (upper panels) and $y=0$ plane (lower panels) in models
CE1 (left) and CE4 (right).
Above each upper panel is the elapsed time $t$ after the cloud begins to collapse, the elapsed time $t_{\rm ps}$ after protostar formation and the protostellar mass $M_{\rm ps}$.
White contours in the lower panels delineate the boundary between the infalling and outflowing gases, within which the gas outflows from the central region.
}
\label{fig:11}
\end{figure}

Figure~\ref{fig:11}  shows the density and velocity distributions when the protostellar mass reaches $M_{\rm ps}=3\msun$ in models CE1 (left panels) and CE4 (right panels). 
It is seen that diffusivity affects the disc size: a larger diffusivity effectively dissipates the magnetic field in the disc, weakening the angular momentum transport by magnetic effects \citep[e.g.][]{masson16,wurster16,tomida15,tsukamoto15}. 
Consequently, a large disc forms. 
As the outflow is driven by the  rotationally supported disc, its foot is located near the disc surface; therefore, the disc size influences the outflow driving region (see lower panels of Fig.~\ref{fig:11}).

Although diffusivity affects magnetic field lines at the small scales around the disc region, if the initial magnetic fields are strong, diffusivity has little impact on the field lines  at large scales around the cloud core and the outflow. 
Instead, the structure of the magnetic field lines at the cloud core scale is primarily determined by the collapse geometry.
Note that, because the amplification of the magnetic field in a small scale determines the outflow properties, the outflow may not be determined by the global collapse geometry when the magnetic field of the prestellar cloud is weak \citep{tomisaka02}. 
In such cases, the degree of magnetic diffusivity may affect large-scale structure of the outflow (see \S\ref{sec:mag} and \ref{sec:appendix}).

Whereas the outflow at large scales significantly contributes to the total outflow momentum, the outflow momentum at small scales has a negligible contribution.
Therefore, because the outflow driving region is located at small scales near the disc region, the difference in diffusivities among the models has a negligible impact on the outflow momentum (Fig.~\ref{fig:10}) when the initial magnetic field is strong. 
In other words, the outflow momentum is determined by the opening angle or geometry of the large-scale magnetic field lines, which is independent of magnetic diffusivity.

\subsection{Dependence of Magnetic Field Strength}
\label{sec:mag}
The fiducial models A--F (Table~\ref{table:1}) are simulated under a strong magnetic field with a mass-to-flux ratio of two. 
In observations of prestellar cores, the magnetic and gravitational energies are comparable and the most plausible mass-to-flux ratio is $\mu = 2-3$ \citep{crutcher99}. 
Recent observations by \citet{tan13} imply that massive starless cores are primary supported by the strong magnetic fields that control the subsequent core accretion phase. 
Although clarifying the origin of the strong magnetic field in prestellar cores is beyond the scope of this study, relating the magnetic field strength to the outflow properties may be important. 
To this end, we prepared four models with weak magnetic fields, namely CW1 and EW1 (with $\mu=5$) and CW2 and EW2 (with $\mu=10$). 
CW and EW have the same cloud parameters as their C and E counterparts, differing only in their magnetic field strengths (Table~\ref{table:1}). 
The cloud evolution of these models is conducted identically to that in the fiducial models.

\begin{figure}
\includegraphics[width=160mm]{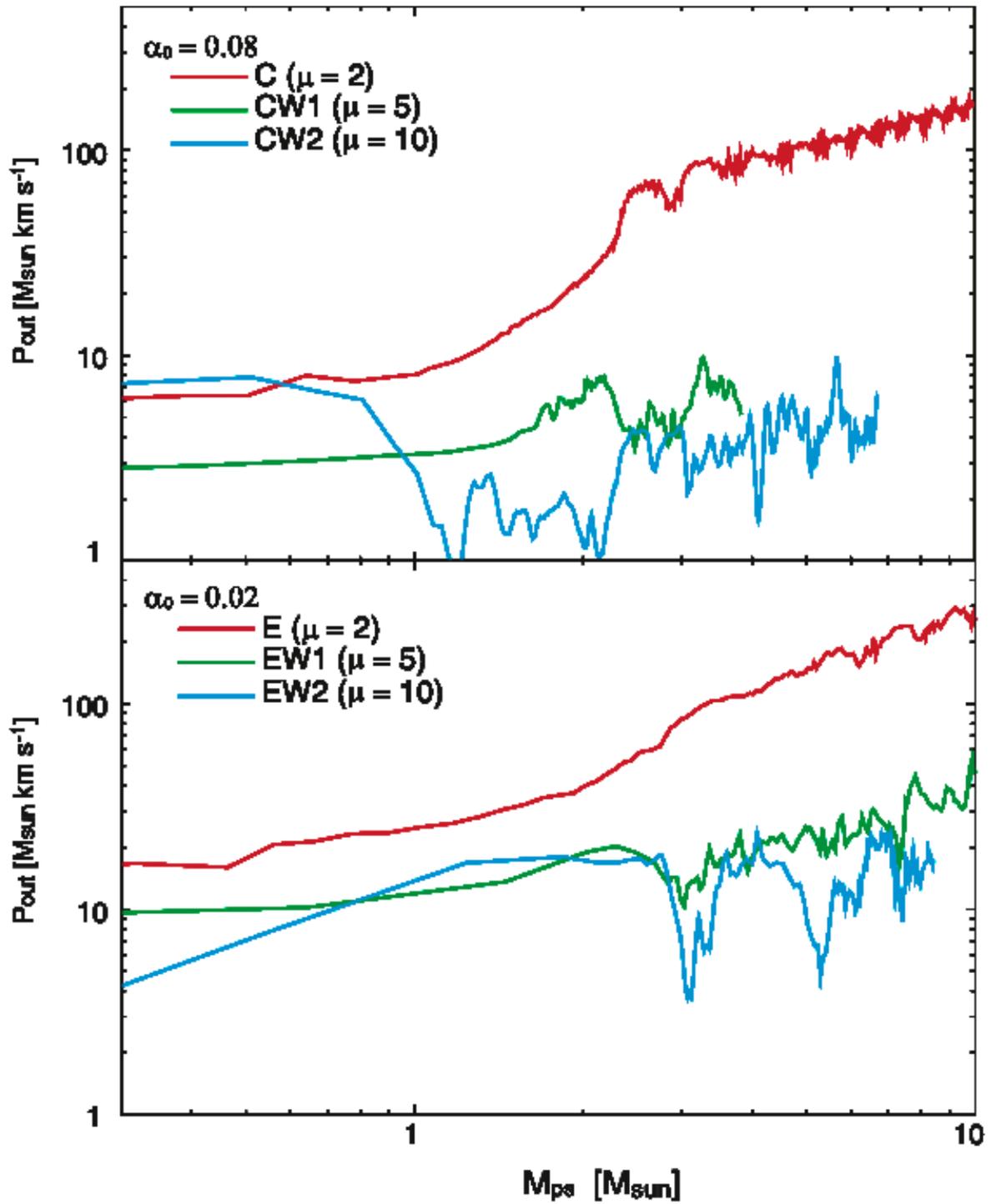}
\caption{
Outflow momenta in models C, CW1 and CW2 (upper panels) and  E, EW1 and EW2 (lower panels) versus protostellar mass.
}
\label{fig:12}
\end{figure}

Figure~\ref{fig:12} shows the outflow momenta in models C (fiducial model), CW1, CW2, E (fiducial model), EW1 and EW2 versus protostellar mass. 
The outflow momenta in the weak (CW1, CW2, EW1 and EW2) and strong (C and E) magnetic field models are not significantly different up to $M_{\rm ps} = 1-2\msun$, but noticeable differences appear beyond this mass.
In the weak magnetic field models, the outflow physical quantities (mass, momentum, kinetic energy and momentum flux) are smaller than those in the strong magnetic field models after protostar formation, especially in the range  $M_{\rm ps} \gtrsim 1-2\msun$. 
When the protostellar mass exceeds $\sim3\msun$, the outflow momenta among the models differs by over one order of magnitude. 
In the weak magnetic field models, the outflow finally disappears. 
Further calculation in the weak magnetic field models leads to recurrent fragmentation.

Figure~\ref{fig:13} shows the time evolution of the weak magnetic field model EW2 ($\mu=10$). 
The outflow appears after protostar formation (Fig.~\ref{fig:13}{\it a}) and reaches $\sim300$\,AU from the central protostar (Fig.~\ref{fig:13}{\it b}). 
Thereafter, the outflow weakens and shrinks (Fig.~\ref{fig:13}{\it c}). Eventually, the outflow component disappears and a torus or toroid-like structure, supported mainly by the magnetic field, appears. 
Conversely, in the strong magnetic field model E, the outflow ultimately reaches $\sim5000$\,AU. 
Thus, changing the initial magnetic field strength qualitatively changes the results.

Our results imply that a magnetic energy comparable to the gravitational energy may be necessary to reproduce the observed outflow properties around a massive protostar. 
However, it should be noted that the necessary magnetic field strength for outflow driving is still controversial.
In the ideal MHD approximation, \citet{hennebelle11} showed that outflow appears even when the initial cloud has an extremely weak magnetic field of $\mu=120$.
On the other hand, recent studies have shown that an outflow does not appear in clouds with $\mu=20$ under ideal MHD calculations \citep{lewis17} and $\mu=10$ under non-ideal MHD calculations \citep{wurster16}. 
Extending our calculations using the parameter of mass-to-flux ratio may produce precisely the magnetic field strength required for the observed massive outflows, but doing so is time-prohibitive;  currently, a single run requires two to four months of wall clock time. 
The dependence of magnetic field strength on outflow driving will be the focus of a future study.
We discuss non-ideal MHD effects on outflow driving in Appendix \S\ref{sec:appendix}.

\subsection{Comparison with Previous Studies} 
\label{sec:comparison}
Several theoretical studies of massive star formation have simulated protostellar MHD outflow in the framework of the core accretion scenario \citep{hennebelle11,commercon11,seifried11,seifried12,peters11,peters12,myers13}. 
In this study, we evolved the cloud over a longer duration in resistive MHD simulations, ignoring the radiative effects. In this subsection, we compare our current work with the previous literature.

As the initial state, \citet{hennebelle11} assumed cores with 100 solar masses and $\alpha_0=0.12$. 
They ran ideal MHD equations, also without radiation effects, and computed the outflow evolution in collapsing clouds for $\sim10^4-10^5$\,yr at spatial resolutions of 2 and 8\,AU. 
They achieved an outflow mass of $\sim\msun$ at the end of their calculation. 
Their initial cloud conditions correspond to our models B ($\alpha_0=0.2$) or  C ($\alpha_0=0.08$), which yielded an outflow mass of $\lesssim 10\msun$ (Figs.~\ref{fig:8}{\it a} and {\it d}). 
As \citet{hennebelle11} did not impose a sink, their protostellar mass was not uniquely determined. 
However, their results closely agree with our simulations at relatively small accretion rates.

Under similar conditions to \citet{hennebelle11}, \citet{commercon11} investigated the evolution of a massive star forming cloud with radiation effects and showed that fragmentation is suppressed in an initially strongly magnetised cloud, as was also observed in our study.
In their calculation, the outflow reached $\sim1000$\,AU with a mass accretion rate of $\sim10^{-4}\mdot$. 
Quantitatively, this result is consistent with our calculations.
 However, their simulation results are difficult to compare against observations because they computed only the earliest phase of massive star formation.

\citet{seifried11} and \citet{seifried12} also investigated massive star formation under the core accretion scenario. 
Their sink radius ($r_{\rm sink}=12$\,AU) and spatial resolution (4.7\,AU) were quite large and neither study appears to have resolved the disc forming region  \citep{machida14a}. 
They included radiative effects but ignored the magnetic dissipation process. 
As the disc forming or magnetic dissipation region determines the outflow properties, it cannot be ascertained whether they correctly  (or quantitatively) estimated these properties in their calculation. 
They evolved the cloud to a protostellar mass of $\sim1-4\msun$, where the mass and momentum of the outflow were $0.1\lesssim M_{\rm out}/(\msun) \lesssim 1.1 $ and  $0.2 \le P_{\rm out}/(\msun\, {\rm km}\, {\rm s}^{-1}) \le 2$, respectively.
Both quantities are substantially smaller than our results and  observations, which may be attributed to their lower spatial resolution and larger sink radius.

\citet{peters11} investigated the collapse of massive clouds in a radiation-MHD simulation. 
Although they also reproduced the outflow in massive star forming regions, the scale of their region was very large, with a sink radius of 590\,AU.
As the outflow driving region was completely embedded in their sink, we do not quantitatively compare their results with ours.  
However, the suppression of fragmentation by the magnetic effects was confirmed in both studies.

Our study serves as an advancement on previous studies in two major aspect:
\begin{itemize}
\item We managed to compare the simulated outflow properties with observations using long-term time integration simulations that resolved the regions of magnetic dissipation and outflow driving.  
\item In addition, by varying the mass accretion rates or initial cloud stabilities, we could relate massive star formation dynamics to those of low \citep{machida13} and intermediate star formation. 
\end{itemize}
As described above, this study focused not merely on massive star formation but also covered a wide mass range. 
Moreover, the outflow properties simulated at higher mass accretion rates agreed with those of previous studies. 
Therefore, the results of this study might provide important clues to the better understanding of the massive star formation process.

\subsection{Effects of Radiative Feedback}
\label{sec:feedback}
As described in \S\ref{sec:model}, this study ignored radiative feedback from the protostar, because our goal was simply to relate mass accretion (inflow) to mass outflow driven by magnetic effects. 
However, we compared our results with observations of massive outflows from massive protostars that are very luminous in the late accretion phase.
Based on this comparison, in this subsection, we discuss the effects of radiative feedback from massive protostars.

In recent years, massive star formation has been studied using radiation-hydrodynamic simulations.
\citet{krumholz07} investigated massive star formation through the use of radiation hydrodynamic simulation in which they primary focused on disc fragmentation.
They showed that fragmentation is suppressed around massive protostars because such stars heat their surroundings \citep{toomre64}. 
Although the magnetic field can also suppress fragmentation, they ignored magnetic effects.
Thus, \citet{krumholz07} produced results that are qualitatively close to those of this study, in which fragmentation is suppressed by either radiative or magnetic effects and only a one or a small number of protostars tend to appear around the centres of massive collapsing clouds. 
In our calculations, fragmentation occurs only in model F, which has the highest accretion rate and the most massive initial prestellar cloud core.
When the radiative effects are included, the disc heating may further suppress fragmentation, even for model F.

In addition to fragmentation, the disc heating is also expected to influence the magnetic dissipation process. 
The ionisation rate increases with the disc temperature, causing magnetic dissipation to become ineffective \citep[e.g.][]{machida07}.
In such cases,  a more powerful magnetic wind may be driven from the circumstellar disc \citep{tomisaka02}.
On the other hand, a strong magnetic field may suppress disc growth through magnetic braking even when the mass accretion rate is significantly high. 
When the circumstellar disc does not grow sufficiently, a powerful magnetic outflow may not appear \citep{li11}.   
Therefore, radiative feedback (or disc heating) is expected to have conflicting effects on magnetically driven outflow.
Thus, further investigation into the massive star formation process in the context of the magnetic field and its dissipation process and radiation feedback--especially after the protostar has grown sufficiently---will be necessary to develop a better quantitative comparison between simulation results and observations (see below).

We next discuss the effect of radiative feedback on mass accretion. 
Mass accretion onto the circumstellar disc is necessary in driving the magnetic outflow because the outflow is powered by the release of the gravitational energy of accreting matter.
Radiative feedback from massive protostars is often discussed with regard to whether the mass accretion lasts until the formation of a very massive star  \citep[$\gtrsim50-100\msun$;][]{wolfire87}. 
However, as very massive star formation is beyond the scope of this study, we discuss the protostellar mass at which  mass accretion and circumstellar environment are affected by radiative feedback.  
\citet{krumholz07} showed that the radiation from the protostar does not significantly interrupt mass accretion until the protostellar mass exceeds $\sim20\msun$. 
\citet{kuiper10,kuiper15} also showed that radiative feedback begins to influence the mass accretion and the circumstellar region when the protostellar mass exceeds $\gtrsim 15-20\msun$ \citep{kuiper16}. 
Although the final stellar mass at which the mass accretion is stopped  by radiative feedback is still controversial \citep{kuiper10}, previous studies indicate that radiative feedback does not significantly affect mass accretion until the protostellar mass reaches $\sim10-20\msun$. 
\citet{kuiper16} also showed that, after the protostellar mass exceeds $\gtrsim20-30\msun$, the stellar radiation force is comparable to gravity above and below the disc and strongly affects the circumstellar environment \citep[see also][]{yorke02}.

The comparison of our simulation results with observations in \S\ref{sec:comp-obs} was limited to the range $M_{\rm ps } \lesssim 20-30\msun$, in which we can conclude that radiative feedback does not significantly affect magnetically driven outflow.
On the other hand, the effect of radiative feedback on protostellar outflow cannot be ignored for $M_{\rm ps}\gtrsim20-30\msun$, because the radiative pressure also contributes to the outflow (radiation-driven outflow) from the central region \citep{kuiper16}.

\section{Summary} 
In this study, we investigated the relation between mass accretion onto protostars (or circumstellar disks) and the magnetically driven outflow in gravitationally collapsing cloud cores. 
As an initial state, we assumed a gravitationally unstable prestellar cloud core with spherical symmetry. 
We found that the mass accretion rate onto the central region of a collapsing cloud core is governed by the ratio of the thermal to gravitational energy in the initial cloud, which can be parametrised as $\alpha_0$.
To investigate the effects of the mass accretion rate on outflow driving, we varied $\alpha_0$ and calculated the cloud evolution and outflow propagation.

When the initial prestellar cloud has a strong magnetic field, the outflow is powerful at any mass accretion rate. 
The outflow is driven by the outer disc region, where the magnetic field and neutral gas are well coupled. 
In our simulations, the mass outflow rate was found to be proportional to the mass accretion rate. 
In addition, the ratio of the outflow to the accretion rate does not significantly depend on the mass inflow rate or the initial cloud stability.

Approximately 20--40\% of the infalling gas is ejected by the protostellar outflow from the circumstellar region. 
Regardless of the accretion rate, the outflow has a very wide opening angle at the disc scale and a well-collimated structure at large scales. 
When the prestellar cloud has a strong magnetic field, the outflow structure is determined by the large-scale geometry of the magnetic field lines.  
The outflow travels into a dense infalling envelope and sweeps part of the infalling gas through the  wide opening angle. 
On the cloud core scale, the mass ratio of the outflowing gas (including the gas swept up by the outflow) to the infalling gas exceeds 50\%. 
Thus, over half of the infalling gas is finally ejected into interstellar space by the outflow.

To validate the importance of  magnetically driven outflow, we quantitatively compared our simulated outflow mass, momentum, kinetic energy and momentum flux with their observed counterparts. 
The physical quantities derived from our simulations closely agree with those observed in massive outflows (Fig.~\ref{fig:8}). 
The simulated momentum fluxes are roughly distributed between the maximum and minimum observed momentum fluxes (Fig.~\ref{fig:9}), suggesting that the massive outflows from massive protostars are magnetically driven, as true in low-mass star formation.
On the other hand, the outflow is subdued or absent in cloud cores with an initially weak magnetic field. 
In addition, weakly magnetised clouds fragment multiple times, preventing the formation of a single massive star. 
Because a strong magnetic field sufficiently explains the massive outflow and suppressed vigorous fragmentation, it may be a requirement for massive star formation.

By combining the results of this study with those of \citet{machida13}, we can state that both low- and high-mass stars form through a common fundamental mechanism (i.e. the core accretion scenario). 
Once a protostar has formed in a gravitationally collapsing cloud, a circumstellar disc grows and establishes a surrounding outflow powered by gas accretion or the gravitational energy released by infalling gas; consequently, the outflow momentum and energy increase with the accretion rate. 
In this manner, the accretion rate determines both the physical quantities of the outflow and the resultant stellar mass. 
Thus, the massive outflows observed around massive protostars are evidence of these star's evolution according to the core accretion scenario.
This study only investigated the relation between accretion and outflow rate, and the comparisons between our simulated and previously observed massive outflows are rather tentative. 
In particular, we ignored turbulence, radiation effects and the surrounding environment outside the star forming cloud, all of which will affect massive star formation. 
These issues were not discussed here in detail because doing so would involve broadening our investigation beyond massive star formation. 
However, to more fully understand the formation processes of low- to high-mass stars, it will ultimately be necessary to consider these effects.

\section*{Acknowledgements}
This study has benefited greatly from discussions with  ~T. Hirota, ~K. Tomida, ~K. Motogi,  ~M. Sekiya, and ~T. Tsuribe. 
We are very grateful to an anonymous reviewer for a number of useful suggestions and comments. 
The authors would like to thank Enago (www.enago.jp) for the English language review.
This research used computational resources of the HPCI system provided by (Cyber Sciencecenter, Tohoku University; Cybermedia Center, Osaka University, Earth Simulator, JAMSTEC) through the HPCI System Research Project (Project ID:hp150092, hp160079).
This work was supported by Grants-in-Aid from MEXT (25400232, 26103707).

\clearpage
\begin{table}
\begin{center}
\begin{tabular}{c|ccccccccccc|ccccccc} \hline
{\footnotesize Model} & $f$ & $C_{\eta}$ & $M_{\rm cl} $ & $R_{\rm cl}$ & $B_0$ & $\Omega_0$ & \multirow{2}{*}{$\alpha_0$} & \multirow{2}{*}{$\beta_0$} & \multirow{2}{*}{$\gamma_0$} & \multirow{2}{*}{$\mu$}  \\
& & &  {\scriptsize [$\msun$]} & {\scriptsize [pc]} & [$\mu$\,G] & {\scriptsize [$10^{-14}$\, s$^{-1}$]} &  &  &  &   \\
\hline
A & 1.4 &\multirow{6}{*}{1} & 32  & \multirow{6}{*}{0.28} &  23 & 3.3 & 0.5 & \multirow{6}{*}{0.02} & \multirow{6}{*}{0.2} & \multirow{6}{*}{2} \\
B & 3.4 & & 77  & \multirow{4}{*}{0.28} &  56 &                 5.1     & 0.2 &  &  & \\
C & 8.4 & & 192 &                       &  140 &                8.0      & 0.08 &  &  & \\
D & 17  & & 385 &                       &  280 &                11      & 0.04 &  &  & \\
E & 34  & & 771 &                       &  560 &                16      & 0.02 &  &  & \\
F & 67  & & 1542&                       &  1120&                23      & 0.01&  &  & \\

\hline
CE1& \multirow{4}{*}{8.4} &0.01& \multirow{4}{*}{192} &  \multirow{4}{*}{0.28}  &  \multirow{4}{*}{140} & \multirow{4}{*}{8.0}        & \multirow{4}{*}{0.08}     &   \multirow{4}{*}{0.02} & \multirow{4}{*}{0.2} & \multirow{4}{*}{2} \\
CE2&  &0.1 &  &                      &   &                      &  &  &  & \\
CE3&  &10  &  &                      &   &                      &  &  &  & \\
CE4&  &100 &  &                      &   &                      &  &  &  & \\
\hline
CW1& 8.4 &\multirow{4}{*}{1}& 192 & \multirow{4}{*}{0.28} &  56 &    8.0   & 0.08 &  \multirow{4}{*}{0.02} & 0.03 & 5 \\ 
CW2& 8.4 & & 192 &                      &  28  &  8.0 & 0.08 &  & 0.008    & 10 \\
EW1& 34  & & 771 &                      &  224 &  16  & 0.02 &  & 0.03     & 5 \\
EW2& 34  & & 771 &                      &  112 &  16  & 0.02 &  & 0.008    & 10 \\
\hline
\end{tabular}
\end{center}
\caption{
Model parameters.
Column 1 lists model names. 
Columns 2--7 give the density enhancement factor $f$, diffusion factor $C_\eta$, cloud mass $M_{\rm cl}$, cloud radius $R_{\rm cl}$, magnetic field strength $B_0$ and angular velocity $\Omega_0$, respectively. 
Columns 8, 9 and 10 give the ratios of thermal $\alpha_0$, rotational $\beta_0$ and magnetic energy $\gamma_0$, respectively, to the gravitational energy of the initial cloud core.
Column 11 gives the initial mass-to-flux ratio normalised by the critical value $\mu$.
}
\label{table:1} 
\end{table}  
\clearpage

\clearpage
\clearpage
\clearpage

\clearpage
\clearpage
\clearpage
\clearpage

\begin{figure}
\includegraphics[width=160mm]{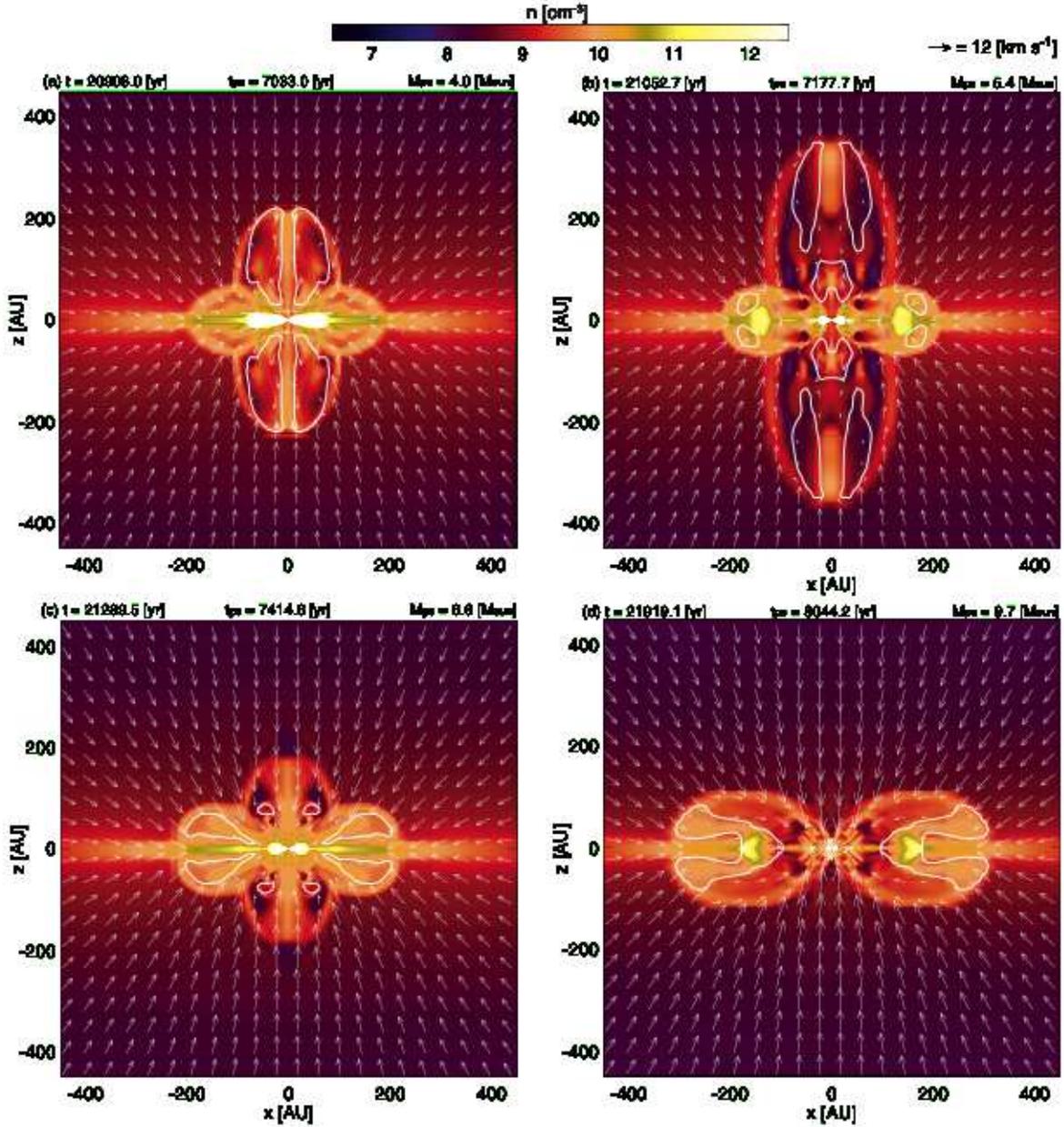}
\caption{
Time sequence of model EW2. 
Each panel plots the density and velocity distributions on the $y=0$ plane.
Above each panel is the elapsed time $t$ after the cloud begins to collapse, the elapsed time $t_{\rm ps}$ after protostar formation and the protostellar mass $M_{\rm ps}$.
Top scales denote density and velocity.
White contours in each panel delineate the boundary between the infalling ($v_r<0$) and outflowing ($v_r>0$) gases, within which the gas outflows from the central region with a radially positive velocity.
}
\label{fig:13}
\end{figure}

\clearpage
\appendix
\section{Difference in Outflow Driving between Ideal and non-ideal MHD models}
\label{sec:appendix}
In \S\ref{sec:mag}, we showed that the outflow driven by the circumstellar disc weakens and disappears quickly in some weakly magnetised models; by contrast, it seems to persist for a long time in previous numerical studies on massive star formation \citep{hennebelle11,seifried12}. 
There are many physical and numerical differences between this and previous studies, including the initial condition, the sink treatment and the spatial resolution. 
We contend that the most important difference is whether magnetic dissipation is taken into account. 
Whereas \citet{hennebelle11} and \citet{seifried12} calculated cloud evolution using ideal MHD equations, we calculated it using non-ideal MHD equations.

\begin{figure}
\includegraphics[width=160mm]{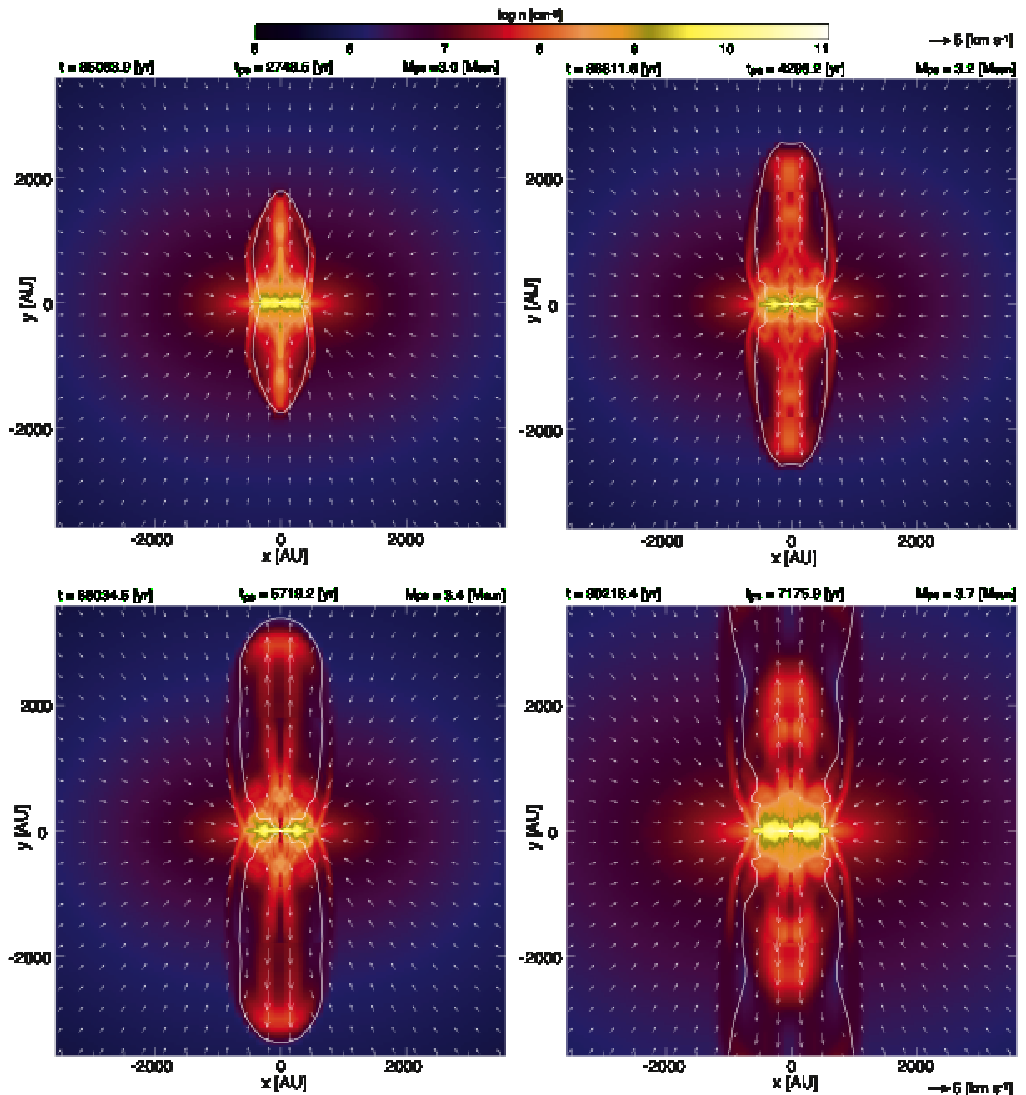}
\caption{
Time sequence of model EW2I (ideal MHD model). 
Each panel plots the density (colour) and velocity (arrows) distributions on the $y=0$ plane.
The elapsed time $t$ after the cloud begins to collapse, the elapsed time $t_{\rm ps}$ after protostar formation and the protostellar mass $M_{\rm ps}$ are listed above each panel. 
The white contours in each panel delineate the boundaries between the infalling ($v_r<0$) and outflowing ($v_r>0$) gases within which the gas outflows from the central region with a radially positive velocity.
}
\label{fig:A1}
\end{figure}

To investigate the effect of the inclusion of magnetic dissipation on outflow driving, we re-calculated the cloud evolution of model EW2 using ideal MHD equations (hereafter we will call the amended model EW2I). 
The numerical settings were identical to those of the models listed in Table~\ref{table:1} except for the inclusion (in EW2, the non-ideal MHD model) or exclusion (in EW2I, the ideal MHD model) of the magnetic field dissipation term. 
As seen from Figure~\ref{fig:13}, in the non-ideal MHD model EW2, after the outflow reaches $\sim400$\,AU, it gradually weakens and finally disappears within $\sim8000$\,yr after protostar formation. 
On the other hand, in the ideal MHD model EW2I (Fig.~\ref{fig:A1}),  outflow reaches $\sim5000$\,AU and does not weaken within $\sim8000$\,yr after protostar formation. 
Thus, magnetic dissipation causes a qualitative difference in the outflow driving process.

The reason for the weakening and disappearance of outflows in the magnetic dissipation models is simple: because the neutral gas is not well coupled to the magnetic field in a magnetically inactive region in which the magnetic field dissipates, the outflow, which is mainly driven from the circumstellar disc by the Lorentz force, does not appear in such regions \citep{machida07}.  
Ohmic resistivity depends on the density (and gas temperature) and a high-density region in the range  $n\lesssim 10^{15}\cm$ has a lower ionisation rate or a larger resistivity \citep[e.g.][]{nakano02}. 
In the massive star formation process, the disc surface density becomes high when the initial magnetic field strength is weak because magnetic braking is not effective even in the magnetically active region where the neutral gas is coupled to the magnetic field. 
Thus, the magnetically inactive region is expected to expand with time and the entire region of the disc might become magnetically inactive. 
In such cases, the outflow does not appear because the magnetic field lines spin free and the magnetic field, which is not coupled to the neutral gas, cannot drive the flow.

\begin{figure}
\includegraphics[width=160mm]{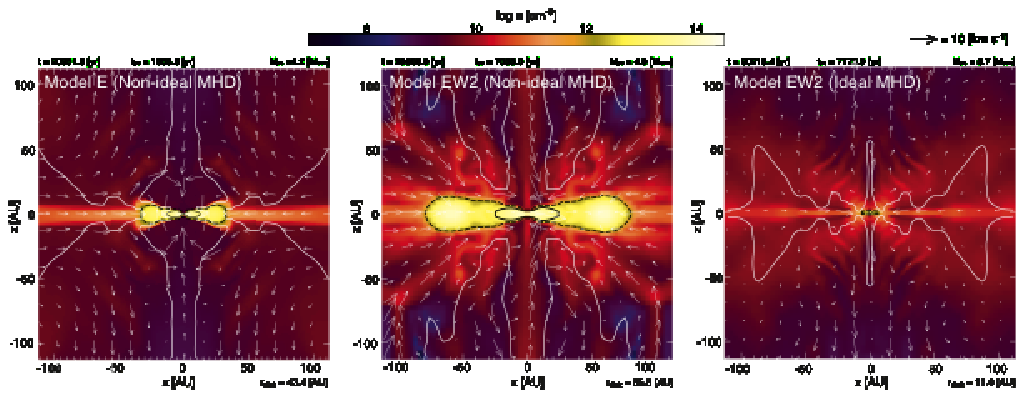}
\caption{
Density (colour) and velocity (arrows) distributions on the $y=0$ plane for models E (left), EW2 (middle) and EW2I (right) when the protostellar mass reaches $\sim 4\msun$. 
White contours in each panel delineate the boundaries between the infalling ($v_r<0$) and outflowing ($v_r>0$) gases within  which the gas outflows from the central region with a radially positive velocity.
The density contours for $n=10^{12}\cm$ (dotted black) and $10^{13}\cm$ (solid black) are plotted in each panel.
The elapsed time $t$ after the calculation begins and that after protostar formation $t_{\rm ps}$ and the protostellar mass $M_{\rm ps}$ are given above each panel. 
The radius of the rotationally supported disc is given below each panel. 
}
\label{fig:A2}
\end{figure}

To compare the size of the disc and the magnetically inactive region, Figure~\ref{fig:A2} plots the density and velocity distributions at density contours of $n=10^{12}$ and $10^{13}\cm$ on the $y=0$ plane for models E (left), EW2 (middle) and EW2I (right) at a protostellar mass of $\sim 4\msun$. 
In the high density region $n>10^{12}\cm$, which corresponds to the region inside the dotted black contour in Figure~\ref{fig:A2}, the magnetic field dissipates through ohmic dissipation.
For model E (Fig.~\ref{fig:A2} left), the outflow is driven by the outer edge of the rotationally supported disc, where the magnetic field is coupled to the neutral gas. 
On the other hand, in model EW2 (Fig.~\ref{fig:A2} middle) the disc and its surrounding envelope have a high density of $\gtrsim10^{12}\cm$, and the overall disc region becomes magnetically inactive.
Although the outflow is just barely driven at this point, it gradually weakens and finally disappears as shown in Figure~\ref{fig:13} because the magnetic field dissipates and gradually weakens in the overall disc region. 
Model EW2I (and EW) has an initially weak magnetic field and a high mass accretion rate.  
The former weakens the magnetic braking while the latter promotes the formation and growth of a massive disc having a high surface density. 
Correspondingly, model EW2 produces a large magnetically inactive region in which outflow is not efficiently driven. 
Even when the magnetic field is weak, the outflow continues to be driven in the ideal MHD model, as shown in Figures~\ref{fig:A1} and  \ref{fig:A2} (right panel) because the magnetic field continues to be amplified by the disc rotation in the ideal MHD approximation.

Finally, to confirm outflow driving in an initially extremely weak magnetised cloud, we prepared the same initial condition and adopted the same numerical settings as in \citet{hennebelle11} and calculated the cloud evolution using {\it ideal} MHD equations and without a sink (hereafter, we call this model HI). 
The initial cloud has a density distribution of 
\begin{equation}
\rho(r) = \dfrac{\rho_{\rm c}}{(1+(r/r_0)^2)}, 
\end{equation}
where  $\rho_c = 1.4\times10^{-20}\,{\rm g}\cm$ and $r_0=0.22$\,pc. 
The initial cloud has a mass of $100\msun$ and a radius of $1.35$\,pc.  
A uniform magnetic field ($B_0=3\times10^{-7}$\,G) is adopted and adjusted to produce a mass-to-flux ratio $\mu=120$ in the overall cloud. 
Whereas an internal turbulent velocity dispersion is given as the source of cloud rotation in \citet{hennebelle11}, a rigid rotation of $\Omega_0=10^{-14}$\,$\rm{s}^{-1}$ is adopted for model HI.

\begin{figure}
\includegraphics[width=140mm]{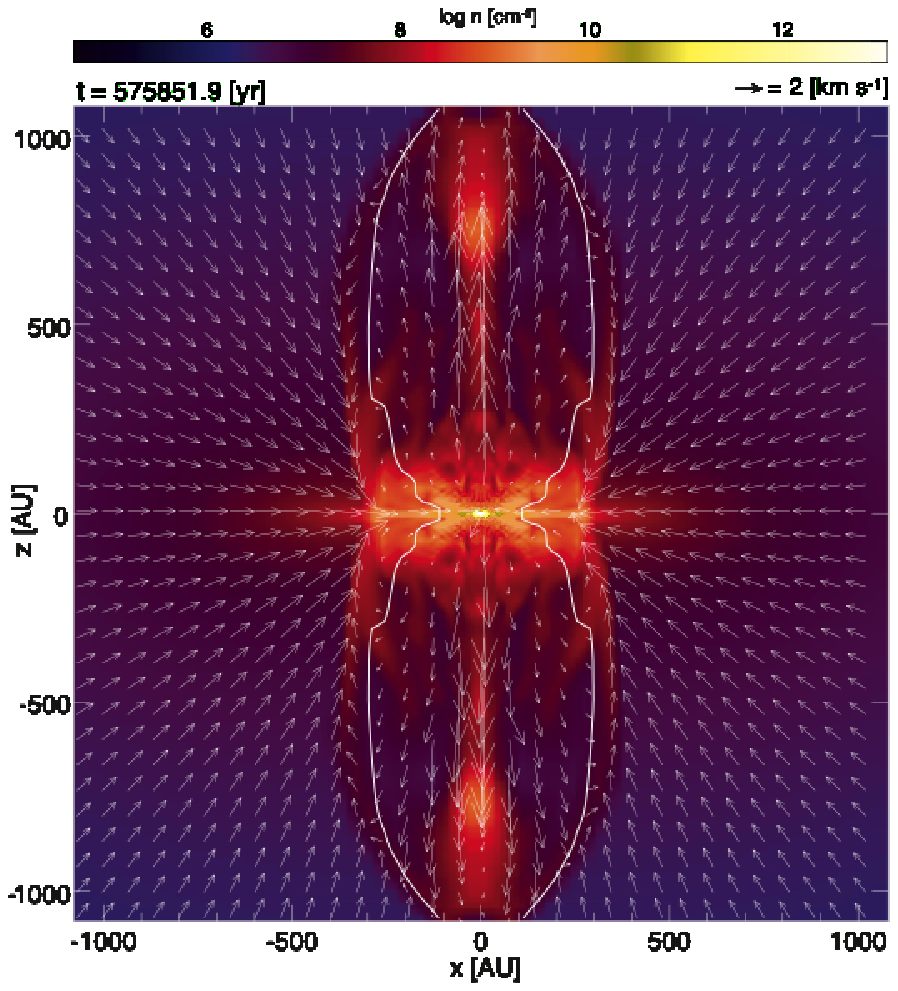}
\caption{
Density (colour) and velocity (arrows) distributions on the $y=0$ plane for model HI. 
The elapsed time after the cloud collapse begins $t$ is given. 
The white contours in each panel delineate the boundaries between the infalling ($v_r<0$) and outflowing ($v_r>0$) gases within  which the gas outflows from the central region with a radially positive velocity.
}
\label{fig:A3}
\end{figure}

Figure~\ref{fig:A3} shows the density and velocity distributions on the $y=0$ plane at $t=5.8\times10^5$\,yr after the cloud begins to collapse. 
Although the initial cloud has an extremely weak magnetic field, the outflow continues and does not weaken by the end of the calculation (see also \S\ref{sec:comparison}). 
This indicates that, in the ideal MHD approximation, the outflow can be driven by the circumstellar disc even when the initial magnetic field is extremely weak. 
Using an ideal MHD simulation, \citet{tomisaka02} also showed that magnetic field lines are highly twisted by the rotational motion of the central region and that the amplified magnetic field drives outflow. 
As we reproduced a long-lived outflow in a cloud with an extremely weak magnetic field using ideal MHD calculations, our result does not contradict previous studies using ideal MHD calculations \citep{tomisaka02,hennebelle11,seifried12}.
On the other hand, in the non-ideal MHD model, since the magnetic field dissipates when the density of the disc and the surrounding medium is high, the magnetic field lines spin free and the magnetic field is not amplified. 
Therefore, even when outflow appears just after protostar or disc formation, it can occasionally weaken or disappear for  short periods of time, as shown in Figure~\ref{fig:13}.

Observations of star forming clouds indicate that their magnetic fields are strong with magnetic energies comparable to their gravitational energies  \citep{Crutcher10}. 
Thus, the assumption of a star forming cloud with an extremely weak magnetic field is not very realistic.
This study focused only on the relation between the mass accretion and ejection rates, and clarifying the outflow driving condition in high-mass star formation is beyond its scope. 
In addition, much more computationally expensive calculation would be required to accurately determine the outflow driving conditions at different magnetic field strengths.
However, as it is important to comprehensively understand the outflow driving condition in the high-mass star formation process, we will focus on this issue in future work.

\clearpage

\begin{thebibliography}{}{}
\bibitem[Arce et al.(2007)]{arce07} 
Arce, H.~G., Shepherd, D., Gueth, F., et al.\ 2007, Protostars and Planets V, 245 

\bibitem[Arce et al.(2010)]{arce10} 
Arce, H.~G., Borkin, M.~A., Goodman, A.~A., Pineda, J.~E., \& Halle, M.~W.\ 2010, ApJ, 715, 1170 

\bibitem[Alves et al.(2007)]{alves07} 
Alves, J., Lombardi, M., \& Lada, C.~J.\ 2007, A\&A, 462, L17 

\bibitem[Andr{\'e} et al.(2010)]{andre10} 
Andr{\'e}, P., Men'shchikov, A., Bontemps, S., et al.\ 2010, A\&A, 518, L102 

\bibitem[Bate(1998)]{bate98} 
Bate, M.~R.\ 1998, ApJL, 508, L95 

\bibitem[Bate(2011)]{bate11} 
Bate, M.~R.\ 2011, MNRAS, 417, 2036 

\bibitem[Bate et al.(2014)]{bate14} 
Bate, M.~R., Tricco, T.~S., \& Price, D.~J.\ 2014, MNRAS, 437, 77 

\bibitem[Beltr{\'a}n et al.(2011)]{beltran11} 
Beltr{\'a}n, M.~T., Cesaroni, R., Zhang, Q., et al.\ 2011, A\&A, 532, A91 

\bibitem[Bally \& Lada(1983)]{bally83} 
Bally, J., \& Lada, C.~J.\ 1983, ApJ, 265, 824 

\bibitem[Bergin \& Tafalla(2007)]{bergin07} 
Bergin, E.~A., \& Tafalla, M.\ 2007, ARA\&A, 45, 339 

\bibitem[Beuther et al.(2002)]{beuther02} 
Beuther, H., Schilke, P., Sridharan, T.~K., et al.\ 2002, A\&A, 383, 892 

\bibitem[Beuther et al.(2007)]{beuther07} 
Beuther, H., Churchwell, E.~B., McKee, C.~F., \& Tan, J.~C.\ 2007, Protostars and Planets V, 165 

\bibitem[Beuther et al.(2010)]{beuther11} 
Beuther, H., Vlemmings, W.~H.~T., Rao, R., \& van der Tak, F.~F.~S.\ 2010, ApJL, 724, L113 

\bibitem[Bonnor(1956)]{bonnor56}
Bonnor, W. B. 1956, MNRAS, 116, 351

\bibitem[\protect\citeauthoryear{Bontemps et al.}{1996}]{bontemps96} 
Bontemps, S., Andre, P., Terebey, S., \& Cabrit, S.\ 1996, A\&A, 311, 858 

\bibitem[Burkhart et al.(2015)]{burkhart15} 
Burkhart, B., Lazarian, A., Balsara, D., Meyer, C., \& Cho, J.\ 2015, ApJ, 805, 118 

\bibitem[Blandford \& Payne(1982)]{blandford82} 
Blandford, R.~D., \& Payne, D.~G.\ 1982, MNRAS, 199, 883 

\bibitem[Carrasco-Gonz{\'a}lez et al.(2015)]{caarrasco15} 
Carrasco-Gonz{\'a}lez, C., Torrelles, J.~M., Cant{\'o}, J., et al.\ 2015, Science, 348, 114 

\bibitem[Chabrier(2003)]{chabrier03} 
Chabrier, G.\ 2003, PASP, 115, 763 

\bibitem[Cabrit \& Bertout(1990)]{cabrit90} 
Cabrit, S., \& Bertout, C.\ 1990, ApJ, 348, 530 

\bibitem[Cabrit \& Bertout(1992)]{cabrit92} 
Cabrit, S., \& Bertout, C.\ 1992, A\&A, 261, 274 

\bibitem[Commer{\c c}on et al.(2011)]{commercon11} Commer{\c c}on, 
B., Hennebelle, P., \& Henning, T.\ 2011, ApJL, 742, L9 

\bibitem[Crutcher(1999)]{crutcher99} 
Crutcher, R.~M.\ 1999, ApJ, 520, 706 

\bibitem[Crutcher et al.(2010)]{Crutcher10} 
Crutcher, R.~M., Wandelt, B., Heiles, C., Falgarone, E., \& Troland, T.~H.\ 2010, ApJ, 725, 466 

\bibitem[Dapp \& Basu(2010)]{dapp10}
Dapp, W.~B., \& Basu, S.\ 2010, A\&A, 521, L56 

\bibitem[de Villiers et al.(2014)]{villiers14} 
de Villiers, H.~M., Chrysostomou, A., Thompson, M.~A., et al.\ 2014, MNRAS, 444, 566 

\bibitem[de Villiers et al.(2015)]{villiers15} 
de Villiers, H.~M., Chrysostomou, A., Thompson, M.~A., et al.\ 2015, MNRAS, 449, 119 

\bibitem[Ebert(1955)]{ebert55}
 Ebert, R. 1955, Z. Astrophys., 37, 222

\bibitem[Enoch et al.(2006)]{enoch06} 
Enoch, M.~L., Young, K.~E., Glenn, J., et al.\ 2006, ApJ, 638, 293 

\bibitem[Falgarone et al.(2008)]{falgarone08} 
Falgarone, E., Troland, T.~H., Crutcher, R.~M., \& Paubert, G.\ 2008, A\&A, 487, 247 

\bibitem[Frank et al.(2014)]{frank14} 
Frank, A., Ray, T.~P., Cabrit, S., et al.\ 2014, Protostars and Planets VI, 451 

\bibitem[Girart et al.(2009)]{girart09} 
Girart, J.~M., Beltr{\'a}n, M.~T., Zhang, Q., Rao, R., \& Estalella, R.\ 2009, Science, 324, 1408 

\bibitem[Hatchell et al.(2007)]{hatchell07} 
Hatchell, J., Fuller, G.~A., \& Richer, J.~S.\ 2007, A\&A, 472, 187 

\bibitem[Hennebelle \& Fromang(2008)]{hennebelle08} 
Hennebelle, P., \& Fromang, S.\ 2008, A\&A, 477, 9 

\bibitem[Hennebelle et al.(2011)]{hennebelle11} 
Hennebelle, P., Commer{\c c}on, B., Joos, M., et al.\ 2011, A\&A, 528, A72 

\bibitem[Joos et al.(2012)]{joos12} 
Joos, M., Hennebelle, P., \& Ciardi, A.\ 2012, A\&A, 543, A128 

\bibitem[Joos et al.(2013)]{joos13} 
Joos, M., Hennebelle, P., Ciardi, A., \& Fromang, S.\ 2013, A\&A, 554, AA17 

\bibitem[Konigl(1989)]{konigl89} 
Konigl, A.\ 1989, ApJ, 342, 208 

\bibitem[Konigl \& Pudritz(2000)]{konigl00} 
Konigl, A., \& Pudritz, R.~E.\ 2000, Protostars and Planets IV, 759 

\bibitem[K{\"o}nyves et al.(2010)]{konyves10} 
K{\"o}nyves, V., Andr{\'e}, P., Men'shchikov, A., et al.\ 2010, A\&A, 518, L106 

\bibitem[Kroupa(2002)]{kroupa02} 
Kroupa, P.\ 2002, Science, 295, 82 

\bibitem[Krumholz et al.(2007)]{krumholz07} 
Krumholz, M.~R., Klein, R.~I., \& McKee, C.~F.\ 2007, ApJ, 656, 959 

\bibitem[Krumholz et al.(2009)]{krumholz09} 
Krumholz, M.~R., Klein, R.~I., McKee, C.~F., Offner, S.~S.~R., \& Cunningham, A.~J.\ 2009, Science, 323, 754 

\bibitem[Kuiper et al.(2010)]{kuiper10} 
Kuiper, R., Klahr, H., Beuther, H., \& Henning, T.\ 2010, ApJ, 722, 1556 

\bibitem[Kuiper et al.(2015)]{kuiper15} 
Kuiper, R., Yorke, H.~W., \& Turner, N.~J.\ 2015, ApJ, 800, 86 

\bibitem[Kuiper et al.(2016)]{kuiper16} 
Kuiper, R., Turner, N.~J., \& Yorke, H.~W.\ 2016, arXiv:1609.05208 

\bibitem[Larson(1969)]{larson69} 
Larson, R. B., 1969, MNRAS, 145, 271.

\bibitem[Lewis \& Bate(2017)]{lewis17} 
Lewis, B.~T., \& Bate, M.~R.\ 2017, arXiv:1701.08741 

\bibitem[Li et al.(2015)]{li15} 
Li, H.-B., Yuen, K.~H., Otto, F., et al.\ 2015, Nature, 520, 518 

\bibitem[Li et al.(2011)]{li11} 
Li, Z.-Y., Krasnopolsky, R., \& Shang, H.\ 2011, ApJ, 738, 180 

\bibitem[L{\'o}pez-Sepulcre et al.(2009)]{lopes09} 
L{\'o}pez-Sepulcre, A., Codella, C., Cesaroni, R., Marcelino, N., \& Walmsley, C.~M.\ 2009, A\&A, 499, 811 

\bibitem[Lynden-Bell(2003)]{lynden-Bell03} 
Lynden-Bell, D.\ 2003, MNRAS, 341, 1360 

\bibitem[Troland \& Crutcher(2008)]{troland08} 
Troland, T.~H., \& Crutcher, R.~M.\ 2008, ApJ, 680, 457-465 

\bibitem[Masson et al.(2016)]{masson16} 
Masson, J., Chabrier, G., Hennebelle, P., Vaytet, N., \& Commer{\c c}on, B.\ 2016, A\&A, 587, A32 

\bibitem[Maud et al.(2015)]{maud15} 
Maud, L.~T., Moore, T.~J.~T., Lumsden, S.~L., et al.\ 2015, MNRAS, 453, 645 

\bibitem[Matzner \& McKee(2000)]{matzner00} 
Matzner, C.~D., \& McKee, C.~F.\ 2000, ApJ, 545, 364 

\bibitem[\protect\citeauthoryear{Machida et al.}{2004}]{machida04} 
 Machida, M. N., Tomisaka, K., \& Matsumoto, T.\ 2004, MNRAS, 348, L1 

\bibitem[\protect\citeauthoryear{Machida \etal}{2005a}]{machida05a}
 Machida, M. N., Matsumoto, T., Tomisaka, K., \& Hanawa, T. 2005, MNRAS, 362, 369

\bibitem[\protect\citeauthoryear{Machida \etal}{2005b}]{machida05b} 
 Machida, M. N., Matsumoto, T., Hanawa, T., \& Tomisaka, K. 2005b, MNRAS, 362, 382 

\bibitem[\protect\citeauthoryear{Machida et al.}{2007}]{machida07} 
 Machida, M.~N., Inutsuka, S., \& Matsumoto, T.\ 2007, ApJ, 670, 1198 

\bibitem[Machida et al.(2008)]{machida08} 
 Machida, M.~N., Inutsuka, S., \& Matsumoto, T.\ 2008, ApJ, 676, 1088

\bibitem[\protect\citeauthoryear{Machida et al.}{2010a}]{machida10} 
 Machida, M.~N., Inutsuka, S., \& Matsumoto, T.\ 2010, ApJ, 724, 1006 

\bibitem[\protect\citeauthoryear{Machida et al.}{2011a}]{machida11a} 
 Machida, M.~N., Inutsuka, S., \& Matsumoto, T.\ 2011a, ApJ, 729, 42 

\bibitem[Machida et al.(2011b)]{machida11b} 
Machida, M.~N., Inutsuka, S., \& Matsumoto, T.\ 2011b, PASJ, 63, 555 

\bibitem[Machida \& Matsumoto(2011c)]{machida11c} 
Machida, M.~N., \& Matsumoto, T.\ 2011c, MNRAS, 413, 2767 

\bibitem[Machida \& Matsumoto(2012)]{machida12} 
 Machida, M.~N., \& Matsumoto, T.\ 2012, MNRAS, 421, 588 

\bibitem[Machida \& Hosokawa(2013)]{machida13} 
Machida, M.~N., \& Hosokawa, T.\ 2013, MNRAS, 431, 1719 

\bibitem[Machida et al.(2014)]{machida14a} 
Machida, M.~N., Inutsuka, S., \& Matsumoto, T.\ 2014, MNRAS, 438, 2278 

\bibitem[Machida(2014)]{machida14} 
Machida, M.~N.\ 2014, ApJL, 796, L17 

\bibitem[Machida \& Nakamura(2015)]{machida15} 
Machida, M.~N., \& Nakamura, T.\ 2015, MNRAS, 448, 1405 

\bibitem[Matsumoto \& Hanawa(2003)]{matsu03} 
Matsumoto, T., \& Hanawa, T.\ 20033, ApJ, 595, 913 

\bibitem[Masunaga \& Inutsuka(2000)]{masunaga00} 
Masunaga, H., \& Inutsuka, S., 2000, ApJ, 531, 350

\bibitem[Motte et al.(1998)]{motte98} 
Motte, F., Andre, P., \& Neri, R.\ 1998, A\&A, 336, 150 

\bibitem[Myers et al.(2013)]{myers13} 
Myers, A.~T., McKee, C.~F., Cunningham, A.~J., Klein, R.~I., \& Krumholz, M.~R.\ 2013, ApJ, 766, 97 

\bibitem[Nakano \& Umebayashi(1980)]{nakano80} 
Nakano, T., \& Umebayashi, T.\ 1980, PASJ, 32, 613 

\bibitem[Nakano \& Umebayashi(1986)]{nakano86} 
Nakano, T., \& Umebayashi, T.\ 1986, MNRAS, 218, 663 

\bibitem[Nakano et al.(1995)]{nakano95} 
Nakano, T., Hasegawa, T., \& Norman, C.\ 1995, ApJ, 450, 183 

\bibitem[\protect\citeauthoryear{Nakano et al.}{2002}]{nakano02} 
 Nakano, T., Nishi, R., \& Umebayashi, T.\ 2002, ApJ, 573, 199 

\bibitem[Nutter \& Ward-Thompson(2007)]{nutter07} 
Nutter, D., \& Ward-Thompson, D.\ 2007, MNRAS, 374, 1413 

\bibitem[Richer et al.(2000)]{richer00} 
Richer, J.~S., Shepherd, D.~S., Cabrit, S., Bachiller, R., \& Churchwell, E.\ 2000, Protostars and Planets IV, 867 

\bibitem[Ridge \& Moore(2001)]{ridge01} 
Ridge, N.~A., \& Moore, T.~J.~T.\ 2001, A\&A, 378, 495 

\bibitem[Palau et al.(2013)]{palau13} 
Palau, A., S{\'a}nchez Contreras, C., Sahai, R., S{\'a}nchez-Monge, {\'A}., \& Rizzo, J.~R.\ 2013, MNRAS, 428, 1537 

\bibitem[Pudritz et al.(2007)]{pudritz07} 
Pudritz, R.~E., Ouyed, R., Fendt, C., \& Brandenburg, A.\ 2007, Protostars and Planets V, 277 

\bibitem[Peters et al.(2011)]{peters11} 
Peters, T., Banerjee, R., Klessen, R.~S., \& Mac Low, M.-M.\ 2011, ApJ, 729, 72 

\bibitem[Peters et al.(2012)]{peters12} 
Peters, T., Klaassen, P.~D., Mac Low, M.-M., Klessen, R.~S., \& Banerjee, R.\ 2012, ApJ, 760, 91 

\bibitem[Pudritz \& Norman(1983)]{pudritz83} 
Pudritz, R.~E., \& Norman, C.~A.\ 1983, ApJ, 274, 677 

\bibitem[Price et al.(2012)]{price12} 
Price, D.~J., Tricco, T.~S., \& Bate, M.~R.\ 2012, MNRAS, 423, L45 

\bibitem[Pudritz \& Norman(1986)]{pudritz86} 
Pudritz, R.~E., \& Norman, C.~A.\ 1986, ApJ, 301, 571 

\bibitem[Pudritz et al.(2007)]{pudritz07} 
Pudritz, R.~E., Ouyed, R., Fendt, C., \& Brandenburg, A.\ 2007, Protostars and Planets V, 277

\bibitem[Qiu et al.(2009)]{qiu09} 
Qiu, K., Zhang, Q., Wu, J., \& Chen, H.-R.\ 2009, ApJ, 696, 66 

\bibitem[Saigo \& Tomisaka(2006)]{saigo06} 
Saigo, K., \& Tomisaka, K.\ 2006, ApJ, 645, 381 

\bibitem[Saigo et al.(2008)]{saigo08} 
Saigo, K., Tomisaka, K., \& Matsumoto, T.\ 2008, ApJ, 674, 997-1014 

\bibitem[Sakurai et al.(2016)]{sakurai16} 
Sakurai, Y., Vorobyov, E.~I., Hosokawa, T., et al.\ 2016, MNRAS, 459, 1137 

\bibitem[Seifried et al.(2011)]{seifried11} 
Seifried, D., Banerjee, R., Klessen, R.~S., Duffin, D., \& Pudritz, R.~E.\ 2011, MNRAS, 417, 1054 

\bibitem[Seifried et al.(2012)]{seifried12} 
Seifried, D., Pudritz, R.~E., Banerjee, R., Duffin, D., \& Klessen, R.~S.\ 2012, MNRAS, 422, 347 

\bibitem[Tan et al.(2013)]{tan13} 
Tan, J.~C., Kong, S., Butler, M.~J., Caselli, P., \& Fontani, F.\ 2013, ApJ, 779, 96 

\bibitem[Tan et al.(2014)]{tan14} 
Tan, J.~C., Beltr{\'a}n, M.~T., Caselli, P., et al.\ 2014, Protostars and Planets VI, 149

\bibitem[Tan et al.(2016)]{tan16} 
Tan, J.~C., Kong, S., Zhang, Y., et al.\ 2016, ApJL, 821, L3 

\bibitem[Tobin et al.(2012)]{tobin12} 
Tobin, J.~J., Hartmann, L., Chiang, H.-F., et al.\ 2012, Nature, 492, 83 

\bibitem[Tomida et al.(2010)]{tomida10} 
Tomida, K., Machida, M.~N., Saigo, K., Tomisaka, K., \& Matsumoto, T.\ 2010, ApJL, 725, L239-L244 

\bibitem[Tomida et al.(2013)]{tomida13}
Tomida, K., Tomisaka, K., Matsumoto, T., et al.\ 2013, ApJ, 763, 6 

\bibitem[Tomida et al.(2015)]{tomida15} 
Tomida, K., Okuzumi, S., \& Machida, M.~N.\ 2015, ApJ, 801, 117 

\bibitem[Tomida et al.(2017)]{tomida17} 
Tomida, K., Machida, M.~N., Hosokawa, T., Sakurai, Y., \& Lin, C.~H.\ 2017, ApJL, 835, L11 

\bibitem[Tomisaka(1998)]{tomisaka98} 
Tomisaka, K.\ 1998, ApJL, 502, L163 

\bibitem[Tomisaka(2000)]{tomisaka00} 
Tomisaka, K.\ 2000, ApJL, 528, L41

\bibitem[Tomisaka(2002)]{tomisaka02} 
Tomisaka, K.\ 2002, ApJ, 575, 306 

\bibitem[Toomre(1964)]{toomre64} 
 Toomre, A.\ 1964, ApJ, 139, 1217 

\bibitem[\protect\citeauthoryear{Truelove \etal}{1997}]{truelove97}
 Truelove J, K., Klein R. I., McKee C. F., Holliman J. H., Howell L. H., \& Greenough J. A., 1997, ApJ, 489, L179

\bibitem[Tsukamoto \& Machida(2013)]{tsukamoto13} 
Tsukamoto, Y., \& Machida, M.~N.\ 2013, MNRAS, 428, 1321 

\bibitem[Tsukamoto et al.(2015a)]{tsukamoto15a} 
Tsukamoto, Y., Iwasaki, K., Okuzumi, S., Machida, M.~N., \& Inutsuka, S.\ 2015a, ApJL, 810, L26 

\bibitem[Tsukamoto et al.(2015)]{tsukamoto15} 
Tsukamoto, Y., Iwasaki, K., Okuzumi, S., Machida, M.~N., \& Inutsuka, S.\ 2015, MNRAS, 452, 278 

\bibitem[Uchida \& Shibata(1985)]{uchida85} 
Uchida, Y., \& Shibata, K.\ 1985, PASJ, 37, 515 

\bibitem[Vorobyov \& Basu(2006)]{vorobyov06} 
Vorobyov, E.~I., \& Basu, S.\ 2006, ApJ, 650, 956 

\bibitem[Vorobyov \& Basu(2015)]{vorobyov15} 
Vorobyov, E.~I., \& Basu, S.\ 2015, ApJ, 805, 115 

\bibitem[Yorke \& Sonnhalter(2002)]{yorke02} 
Yorke, H.~W., \& Sonnhalter, C.\ 2002, ApJ, 569, 846 

\bibitem[Walch et al.(2009)]{walch09} 
Walch, S., Burkert, A., Whitworth, A., Naab, T., \& Gritschneder, M.\ 2009, MNRAS, 400, 13 

\bibitem[Walch et al.(2012)]{walch12} 
Walch, S., Whitworth, A.~P., \& Girichidis, P.\ 2012, MNRAS, 419, 760 

\bibitem[Wardle \& Koenigl(1993)]{wardle93} 
Wardle, M., \& Koenigl, A.\ 1993, ApJ, 410, 218 

\bibitem[Wolfire \& Cassinelli(1987)]{wolfire87} 
Wolfire, M.~G., \& Cassinelli, J.~P.\ 1987, ApJ, 319, 850 

\bibitem[Wurster et al.(2016)]{wurster16} 
Wurster, J., Price, D.~J., \& Bate, M.~R.\ 2016, MNRAS, 457, 1037 

\bibitem[Wu et al.(2004)]{wu04}
Wu, Y., Wei, Y., Zhao, M., Shi, Y., Yu, W., Qin, S., \& Huang, M. 2004, A\&A, 426, 503

\bibitem[Wu et al.(2005)]{wu05} 
Wu, Y., Zhang, Q., Chen, H., et al.\ 2005, AJ, 129, 330 

\bibitem[Wurster et al.(2016)]{wurster16} 
Wurster, J., Price, D.~J., \& Bate, M.~R.\ 2016, MNRAS, 457, 1037 

\bibitem[Zhang et al.(2005)]{zhang05} 
Zhang, Q., Hunter, T.~R., Brand, J., et al.\ 2005, ApJ, 625, 864 

\bibitem[Zhang et al.(2014)]{zhang14} 
Zhang, Q., Qiu, K., Girart, J.~M., et al.\ 2014, ApJ, 792, 116 

\bibitem[Zinnecker \& Yorke(2007)]{zinnecker07} 
Zinnecker, H., \& Yorke, H.~W.\ 2007, ARA\&A, 45, 481 
\end{thebibliography}
\end{document}